\documentclass[twocolumn,useAMS,usenatbib]{mn2e}

\usepackage{graphicx}
\usepackage{subfigure}
\usepackage{xcolor}
\usepackage{mathtext,bbm,amsmath,amsfonts,amssymb,indentfirst,syntonly,graphicx}
\usepackage{mathtools}
\usepackage{slashbox}
\usepackage[english]{babel}
\usepackage{calc}
\usepackage{tikz}

\def\bc{\begin{center}}
\def\ec{\end{center}}
\def\be{\begin{eqnarray}}
\def\ee{\end{eqnarray}}

\title[GRB spectra and the luminosity correlations]{Effect of GRB spectra on the empirical luminosity correlations and the GRB Hubble diagram}
\author[H.-N. Lin, X. Li and Z. Chang]
        {Hai-Nan Lin$^{1}$\thanks{e-mail: linhainanjyzjcn@163.com.},
        Xin Li$^{1,2}$\thanks{e-mail: lixin1981@cqu.edu.cn.},
        Zhe Chang$^{3}$\thanks{e-mail: changz@ihep.ac.cn.}\\
$^{1}$Department of Physics, Chongqing University, Chongqing 401331, China\\
$^{2}$State Key Laboratory of Theoretical Physics, Institute of Theoretical Physics, Chinese Academy of Sciences, Beijing 100190, China\\
$^{3}$Institute of High Energy Physics, Chinese Academy of Sciences, Beijing 100049, China\\}
\begin{document}

\date{Accepted xxxx; Received xxxx; in original form xxxx}

\pagerange{\pageref{firstpage}--\pageref{lastpage}} \pubyear{2016}

\maketitle

\label{firstpage}

\begin{abstract}
The spectra of gamma-ray bursts (GRBs) in a wide energy range can usually be well described by the Band function, which is a two smoothly jointed power laws cutting at a breaking energy. Below the breaking energy, the Band function reduces to a cut-off power law, while above the breaking energy it is a simple power law. However, for some detectors (such as the Swift-BAT) whose working energy is well below or just near the breaking energy, the observed spectra can be fitted to cut-off power law with enough precision. Besides, since the energy band of Swift-BAT is very narrow, the spectra of most GRBs can be fitted well even using a simple power law. In this paper, with the most up-to-date sample of Swift-BAT GRBs, we study the effect of different spectral models on the empirical luminosity correlations, and further investigate the effect on the reconstruction of GRB Hubble diagram. We mainly focus on two luminosity correlations, i.e., the Amati relation  and Yonetoku relation. We calculate these two luminosity correlations on both the case that the GRB spectra are modeled by Band function and cut-off power law. It is found that both luminosity correlations only moderately depend on the choice of GRB spectra. Monte Carlo simulations show that Amati relation is insensitive to the high-energy power-law index of the Band function. As a result, the GRB Hubble diagram calibrated using luminosity correlations is almost independent on the GRB spectra.
\end{abstract}


\section{Introduction}\label{sec:introduction}

Gamma-ray bursts (GRBs), mainly emitting radiation in the gamma-ray waveband  which lasts a few seconds, and may be followed by the $X$-ray, optical or radio emissions lasting a few days, are the most energetic explosions in the Universe \citep{Piran:1999,Meszaros:2006,Kumar:2015}. The recent decades have seen great successes in understanding the nature of GRBs thanks to the contribution by various space experiments. The isotropic distribution in the sky strongly implies that GRBs happen in the deep universe \citep{Paciesas:1999}. The cosmological origin of GRBs was eventually confirmed starting form 1997 thanks to the combination of accurate localizations by BeppoSAX and redshift measurements from the follow-up observation with the main optical/IR facilities. The furthest GRB observed at present has redshift $\sim 9.4$ \citep{Cucchiara:2011}, and the total isotropic equivalent radiated energy by a GRB is about $10^{48}\sim 10^{55}$ ergs \citep{Kumar:2015}. The durations of GRBs span about six orders of magnitude, from milliseconds for the shortest to thousands of seconds for the longest. It is widely accepted that GRBs can be divided into two classes according to the duration, because the histogram of duration is fitted well by the summation of two Gauss functions separated at around 2 seconds \citep{Kouveliotou:1993}. Therefore, short and long GRBs are classified depending on whether the duration is shorter or longer than 2 seconds. The light curves of GRBs have various type, from smoothly decaying to highly variable with many spikes \citep{Fishman:1995}. The spectrum of a typical GRB can be well fitted by the Band function \citep{Band:1993}, which is a two smoothly jointed power lows cutting at a breaking energy. Below the breaking energy, the Band function reduces to the cut-off power law (CPL), and above the breaking energy, it is the simple power law (PL). However, some detectors such as the Swift-BAT, work in an energy band well below or just near the breaking energy. Therefore, it is enough to fit the spectra just with the simple PL, and the high-energy power-law index of the Band function couldn't be constrained.

Since the first linear correlation between the spectral lag and the isotropic peak luminosity ($\tau_{\rm lag}-L$ relation) was found by \citet{Norris:2000}, several other empirical correlations have been found. These empirical correlations relate the luminosity to the spectral parameters of GRBs. One of the most famous correlation was found by \cite{Amati:2002}. The Amati relation is a linear (in the logarithmic scale) correlation between the isotropic equivalent energy and the peak energy of $\nu F_{\nu}$ spectra ($E_p-E_{\rm iso}$ relation). Later, \citet{Yonetoku:2004} found a similar correlation between the isotropic peak luminosity and the peak energy ($E_p-L$ relation). Other two-parameter luminosity correlations includes: $V-L$ relation \citep{Fenimore:2000}, $E_p-E_{\gamma}$ relation \citep{Ghirlanda:2004b}, $\tau_{\rm RT}-L$ relation \citep{Schaefer:2007}, and so on.  Besides, there are also three-parameter correlations such as $E_{\rm iso}-E_p-t_b$ relation \citep{Liang:2005}, $L-E_p-T_{0.45}$ relation \citep{Firmani:2006b}, $E_{X,\rm iso}-E_{\gamma,\rm iso}-E_p$ relation \citep{Bernardini:2012,Margutti:2013}, $L-E_p-\Gamma_0$ relation \citep{Liang:2015dua}. Some luminosity correlations are only valid for long GRBs, while others are valid for both short and long GRBs. These correlations are often used to calibrate GRBs as the standard candles and to reconstruct the GRB Hubble diagram \citep{Schaefer:2003,Dai:2004tq,Ghirlanda:2004a,Liang:2005,Firmani:2005,Schaefer:2007,Amati:2008,Basilakos:2008,Liang:2008a,Liang:2008b,Wei:2009, Wei:2010,Liu:2014,Wang:2015cya,Lin:2015b}.

The above luminosity correlations involve three quantities, i.e., the isotropic equivalent energy $E_{\rm iso}$, the isotropic peak luminosity $L$ and the spectral peak energy $E_p$. The derivation of the first two quantities involves the integration of spectra over the ``\,bolometric" $1-10000$ keV energy band. However, most detectors such as Swift-BAT only measure photons in a narrow energy band. Therefore, we should extrapolate the observed spectra to the whole energy band. Most previous works calculated $E_{\rm iso}$ and $L$ by assuming that the spectra are modeled by the Band function. Due to the narrow energy width of detectors, CPL (or even PL) is enough to fit the observed spectra, and the high-energy index ($\beta$) of the Band function can't be constrained. \citet{Schaefer:2007} fixed $\beta$ to be $-2.2$ for all the GRBs that have no measured $\beta$. In fact, about $60\%$ ($41$ out of $69$) GRBs compiled in \citet{Schaefer:2007} have no measurement of $\beta$. The following works \citep{Liang:2008a,Basilakos:2008,Xiao:2009,Wei:2010,Wang:2011,Liu:2014} which cited the data from \citet{Schaefer:2007} also have the underling assumption that the unmeasured $\beta$'s are all set to be $-2.2$. It is especially necessary to check such a setting has some influence on the final luminosity correlations. Using 29 Swift GRBs, \citet{Cabrera:2007} found that the zero-point of Amati relation calculated from CPL is slightly larger than that from Band function, while the intercept does not change significantly. The small change of zero-point of Amati relation may have some influences on the calibration of GRBs.

In this paper, we investigate if the luminosity correlations are affected by the choice of GRB spectra. We search the most up-to-date Swift data archive\footnote{http://swift.gsfc.nasa.gov/.} and find that among $\sim 300$ GRBs which have redshift measurement, only 44 GRBs can be well fitted by CPL so have well defined peak energy. Using the 44 GRBs, we study the $E_p-E_{\rm iso}$ correlation and $E_p-L$ correlation. All the 44 GRBs are fitted by CPL so the high-energy index couldn't be determined. Therefore, we calculate $E_{\rm iso}$ and $L$ by using the CPL instead of the Band function. To comparison, we also calculate $E_{\rm iso}$ and $L$ using the Band function, with the high-energy index fixed to be $-2.2$, the median value of BATSE GRBs \citep{Preece:2000}, as was done by \citet{Schaefer:2007}. We further calibrate the distance of GRBs using these two luminosity correlations respectively, and reconstruct GRB Hubble diagram. We investigate the influence of different spectra on the luminosity correlations and on the reconstruction of GRB Hubble diagram. Throughout this paper, we assume the concordance cosmological model with the fiducial parameters $H_0=70~{\rm km}~{\rm s}^{-1}~{\rm Mpc}^{-1}$, $\Omega_M=0.28$ and $\Omega_{\Lambda}=0.72$.

The rest of this paper is arranged as follows: In Section \ref{sec:observation}, we present the observational properties of 44 Swift-BAT GRBs used in our analysis. In Section \ref{sec:correlation}, we investigate the $E_p-E_{\rm iso}$ relation and $E_p-L$ relation by assuming that the GRB spectrum is modeled by Band function and CPL, respectively, and compare their differences. We also implement Monte Carlo simulations to test the sensitivity of luminosity correlations on the high-energy power-law index. In Section \ref{sec:hubble}, we calibrate the distance of GRBs through these two luminosity correlations, to see whether the resulting GRB Hubble diagram would depend on the choice of spectra. Finally, Section \ref{sec:conclusions} is devoted to a short summary.

\section{The observational properties of Swift GRBs}\label{sec:observation}

The Swift satellite, launched on November 20, 2004, is a first-of-its-kind multi-wavelength observatory dedicated to the study of GRB science. There are three instruments onboard Swift working together to observe GRBs and afterglows in the gamma-ray, $X$-ray, ultraviolet, and optical wavebands. The Burst Alert Telescope (BAT, $15-150$ keV), which is a highly sensitive and large field of view gamma-ray detector working in the energy band $15-150$ keV, detects GRBs and accurately determines their positions on the sky. Soon after the triggering of BAT, the X-ray Telescope (XRT, $0.3-10$ keV) and UV/Optical Telescope (UVOT, $170-600$ nm) quickly repoint the field of view towards GRBs and start the afterglow observation. With the rate of about 100 GRBs per year, Swift has observed about one thousand GRBs until September, 2015, and about 300 GRBs have redshift measurement.

\begin{figure}
  \centering
 \includegraphics[width=0.5\textwidth]{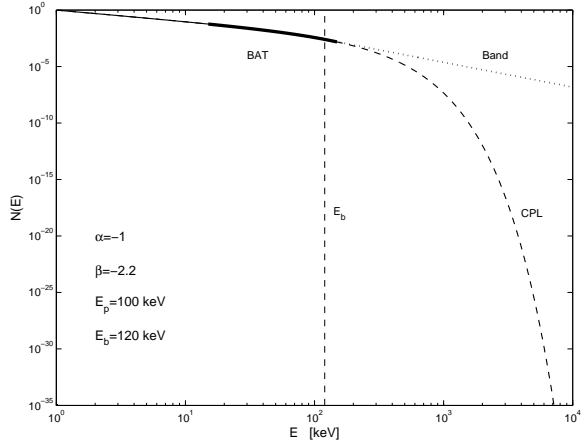}
 \caption{\small{The difference between Band function and CPL. The vertical axis is normalized so that $N(E=1~{\rm keV})=1$. The dashed curve is the CPL, and the dotted curve is the Band function. The vertical dashed line represents the breaking energy $E_b\equiv [(\alpha-\beta)/(2+\alpha)]E_p$. The black solid curve represents the energy band ($15-150$ keV) in which the Swift-BAT is sensitive. The parameters in this figure are: $\alpha=-1$, $\beta=-2.2$, $E_p=100$ keV, and $E_b=120$ keV.}}\label{fig:band}
\end{figure}

For most GRBs, the spectra can be well fitted by a two smoothly jointed power laws cutting at a breaking energy $E_b$, which is well known as the Band function \citep{Band:1993},
\begin{equation}
  N_{\rm Band}(E)=\begin{dcases}
    AE^{\alpha}\exp\left[-(2+\alpha)\frac{E}{E_p}\right] \quad & E\leq \frac{\alpha-\beta}{2+\alpha}E_p\,,\\
    BE^{\beta} \quad & E>\frac{\alpha-\beta}{2+\alpha}E_p\,,
  \end{dcases}
\end{equation}
where $\alpha$ and $\beta$ are the low-energy and high-energy power-law index, respectively, and they vary from bursts to bursts. Statistically,  $\alpha$ peaks around $-1$, and $\beta$ peaks around $-2.2$ \citep{Preece:2000}. The Band function is continuous and smooth at the breaking energy $E_b\equiv(\alpha-\beta)/(2+\alpha)E_p$, and it is monotonously decreasing in the whole energy range. However, it is easy to prove that the $\nu F_{\nu}\propto E^2N(E)$ spectrum has a peak at $E=E_p$, for any $\alpha\geq -2$ and $\beta< -2$. The distribution of $E_p$ depends on the specific detectors. For the BATSE catalog, $E_p$ peaks around $150$ keV \citep{Preece:2000}, while for the Swift-BAT catalog, $E_p$ is relatively lower \citep{Sakamoto:2011}. The Band function was first proposed from analyzing the spectra of GRBs observed by BATSE, which is a gamma-ray detector sensitive in a wide energy band $20-2000$ keV. However, the Swift-BAT is only sensitive in a low and narrow energy band, which is just near or even below the typical breaking energy. Therefore, the observed spectra by Swift-BAT can be fitted using the low-energy end of Band function, i.e., the CPL,
\begin{equation}
  N_{\rm CPL}(E)=AE^{\alpha}\exp\left[-(2+\alpha)\frac{E}{E_p}\right].
\end{equation}
This is to say, the high-energy index $\beta$ can't be well constrained. The CPL is a special case of Band function in the limit $\beta\rightarrow \infty$. Figure \ref{fig:band} shows the difference between the Band function and CPL. CPL falls much more dramatically than Band function above $E_b$.
On the other hand, since the Swift-BAT working energy band is very narrow, for majority of GRBs we can even fit the spectra just with a simple PL, i.e.,
\begin{equation}
  N_{\rm PL}(E)=AE^{\alpha}.
\end{equation}
The $\nu F_{\nu}$ spectrum of CPL also peaks at $E_p$, while PL is monotonous and has no peak.

\begin{figure*}
\centering
 \includegraphics[width=0.8\textwidth]{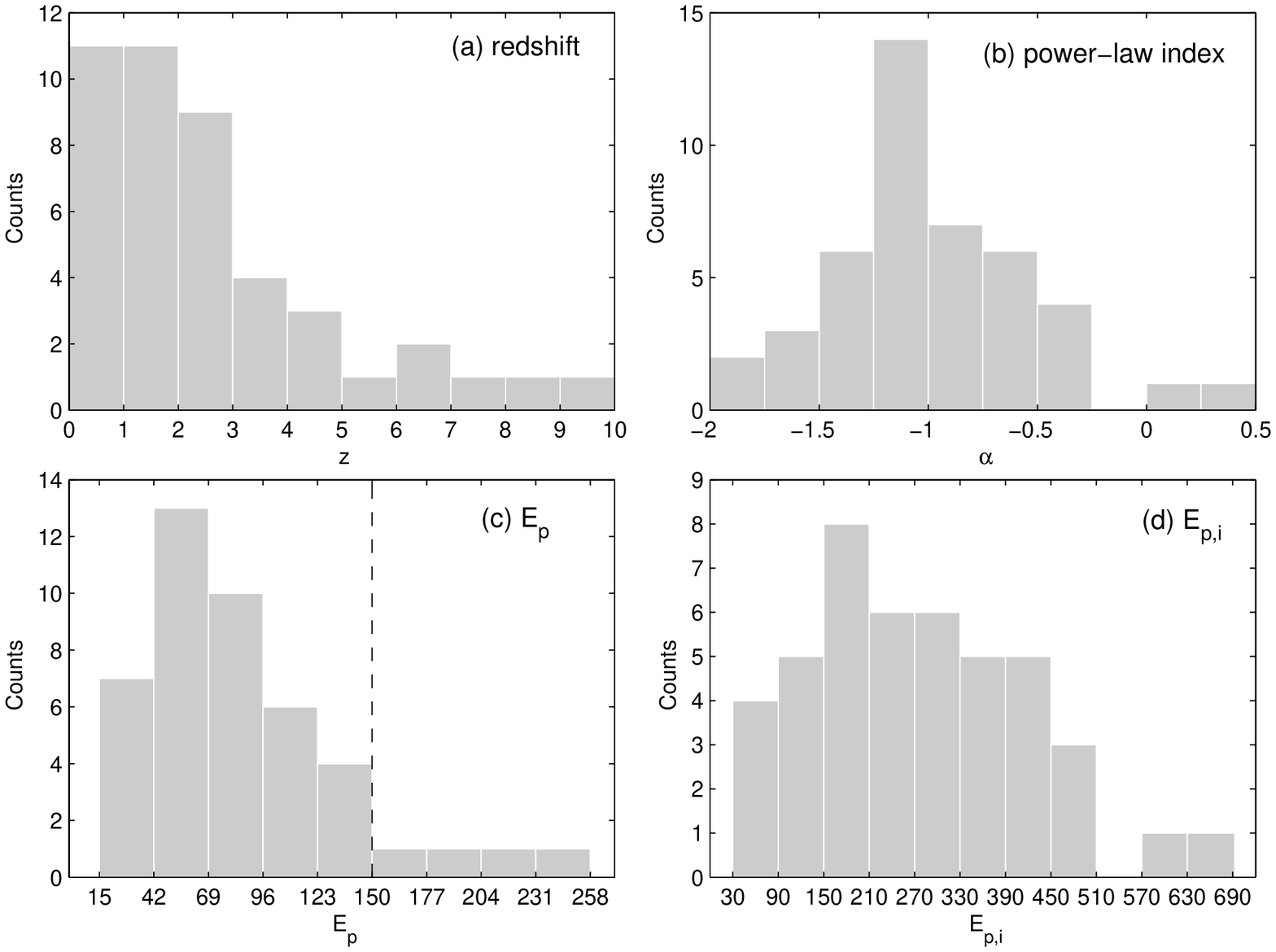}
 \caption{\small{The histograms of (a) the redshift $z$, (b) the power-law index $\alpha$, (c) the observed peak energy $E_p$, and (d) the rest-frame peak energy $E_{p,i}$. The vertical dashed line in panel (c) is the upper boundary of the BAT energy band.}}\label{fig:histogram}
\end{figure*}

Among $\sim 300$ Swift GRBs with measured redshift, only 44 GRBs have a reasonable CPL fit so have well determined peak energy, while the remaining GRBs can be fitted well by the simple PL. Only these 44 GRBs are available to analyze the luminosity correlations. We list the observational properties of the 44 GRBs in Table \ref{tab:parameters1}.
\begin{table*}
\centering
\caption{\small{The observed properties of 44 Swift-BAT GRBs. Column (1): the GRB identifier. Columns (2) and (3): the Right Ascension and Declination (J2000) in unit of degree. Column (4): the GRB duration in which time 90\% photons are recorded. Column (5): the redshift. Column (6): the observed photon fluence in $15-150$ keV energy band in unit of $10^{-7}$ erg cm$^{-2}$. Column (7): the observed peak photon flux in $15-150$ keV energy band in unit of photons cm$^{-2}$ s$^{-1}$. Column (8): the photon index. Column (9): the observed peak energy in unit of keV. Column (10): the GCN Circular numbers (http://gcn.gsfc.nasa.gov/gcn3\b{\,\,}archive.html). All the errors in this table are of 90\% C.L. Note that in the Swift data archive, the uncertainties of $E_p$ (and/or $S$, $P$, $\alpha$) of some GRBs are not symmetric. We symmetrize them by taking the average, i.e., $\sigma=(\sigma_{-}+\sigma_{+})/2$. }}\label{tab:parameters1}
\begin{tabular}{llllllllll}
\hline
(1)	&	(2)	&	(3)	&	(4)	&	(5)	&	(6)	&	(7)	&	(8)	&	(9)	&	(10)	\\
GRBs	&	RA	&	DEC	&	$T_{90}$	&	$z$	&	$S$	&	$P$ 	&	$\alpha$	&	$E_p$ 	&	GCN Circ.	\\
\hline
141220A	&	195.058 &	$+32.146$ &	7.21	&	1.3195	&	$26\pm 1$	&	$8.9\pm 0.7$	&	$-0.62\pm 0.38$	&	$117.4\pm 45.1$	&	 17196;  17202	\\
140518A	&	227.231 &	$+42.396$ &	60.5	&	4.707	&	$10\pm 1$	&	$1.0\pm 0.1$	&	$-0.92\pm 0.61$	&	$43.9\pm 7.6$	&	 16298;  16306	\\
140515A	&	186.071 &	$+15.099$ &	23.4	&	6.32	&	$5.9\pm 0.6$	&	$0.9\pm 0.1$	&	$-0.98\pm 0.64$	&	$51.3\pm 14.7$	&	16267;  16284 \\
140206A	&	145.321 &	$+66.762$ &	93.6	&	2.73	&	$160\pm 3$	&	$19.4\pm 0.5$	&	$-1.04\pm 0.15$	&	$100.9\pm 14.5$	&	 15784;  15805	\\
131117A	&	332.354 &	$-31.761$ &	11	&	4.18 &	$2.5\pm 0.4$	&	$0.7\pm 0.1$	&	$+0.40\pm 1.41$	&	$44.0\pm 7.4$	&	 15490;  15499; 15500	\\
130925A	&	41.186 	&	$-26.146$ &	$\cdots$	&	0.347	&	$410\pm 10$	&	$7.3\pm 0.6$	&	$-1.85\pm 0.14$	&	$33.4\pm 20.0$	& 15246;  15257	\\
130701A	&	357.224 &	$+36.100$ &	4.38	&	1.155	&	$44\pm 1$	&	$17.1\pm 0.7$	&	$-0.90\pm 0.21$	&	$89.2\pm 12.4$	&	 14953;  14959	\\
130612A	&	259.771 &	$+16.729$ &	4	&	2.006	&	$2.3\pm 0.5$	&	$1.7\pm 0.3$	&	$+0.14\pm 1.71$	&	$36.4\pm 7.6$	&	 14874;  14916	\\
130420A	&	196.118 &	$+59.421$ &	123.5	&	1.297	&	$71\pm 3$	&	$3.4\pm 0.2$	&	$-1.52\pm 0.25$	&	$33.2\pm 6.8$	&	 14406;  14419	\\
120923A	&	303.781 &	$+6.255$ 	&	27.2	&	8.5	&	$3.2\pm 0.8$	&	$0.6\pm 0.1$	&	$-0.29\pm 1.66$	&	$44.4\pm 10.6$	&	 13796;  13807	\\
120922A	&	234.758 &	$-20.181$ &	173	&	3.1	&	$62\pm 7$	&	$2.0\pm 0.2$	&	$-1.58\pm 0.36$	&	$38.2\pm 0.0$	&	 13793;  13806	\\
120811C	&	199.690 &	$+62.297$ &	26.8	&	2.671	&	$30\pm 3$	&	$4.1\pm 0.2$	&	$-1.40\pm 0.30$	&	$42.9\pm 5.7$	&	 13622;  13634	\\
120802A	&	44.833 	&	$+13.762$ &	50	&	3.796	&	$19\pm 3$	&	$3.0\pm 0.2$	&	$-1.21\pm 0.47$	&	$57.2\pm 19.4$	&	 13555;  13559	\\
120724A	&	245.193 &	$+3.535$ 	&	72.8	&	1.48	&	$6.8\pm 1.1$	&	$0.6\pm 0.2$	&	$-0.53\pm 1.53$	&	$27.6\pm 7.5$	& 13510;  13517	\\
120326A	&	273.906 &	$+69.248$ &	69.6	&	1.798	&	$26\pm 3$	&	$4.6\pm 0.2$	&	$-1.41\pm 0.34$	&	$41.1\pm 6.9$	&	 13105;  13120	\\
110726A	&	286.713 &	$+56.070$ &	5.2	&	1.036	&	$2.2\pm 0.3$	&	$1.0\pm 0.2$	&	$-0.64\pm 0.87$	&	$46.5\pm 11.8$	&	 12196;  12201	\\
110715A	&	237.665 &	$-46.237$ &	13	&	0.82	&	$118\pm 2$	&	$53.9\pm 1.1$	&	$-1.25\pm 0.12$	&	$120\pm 21$	&	 12158;  12160	\\
110503A	&	132.799 &	$+52.211$ &	10	&	1.613	&	$100\pm 4$	&	$1.35\pm 0.06$	&	$-0.88\pm 0.25$	&	$133\pm 54$	&	 11991;  11995	\\
110422A	&	112.057 &	$+75.100$ &	25.9	&	1.77	&	$410\pm 10$	&	$30.7\pm 1.0$	&	$-0.86\pm 0.10$	&	$149.4\pm 18.5$	&	 11957;  11959	\\
100816A	&	351.738 &	$+26.568$ &	2.9	&	0.8034	&	$20\pm 1$	&	$10.9\pm 0.4$	&	$-0.73\pm 0.24$	&	$170.7\pm 79.7$	& 11102;  11111;  11113	\\
091029	&	60.166 	&	$-55.954$ &	39.2	&	2.752	&	$24\pm 1$	&	$1.8\pm 0.1$	&	$-1.46\pm 0.27$	&	$61.4\pm 17.5$	&	 10097;  10103	\\
091018	&	32.191 	&	$-57.546$ &	4.4	&	0.971	&	$14\pm 1$	&	$10.3\pm 0.4$	&	$-1.77\pm 0.24$	&	$19.2\pm 14.5$	&	 10034;  10040	\\
090926B	&	46.310 	&	$-38.997$ &	109.7	&	1.24	&	$73\pm 2$	&	$3.2\pm 0.3$	&	$-0.52\pm 0.24$	&	$78.3\pm 7.0$	&	 9935;  9939	\\
090618	&	294.008 &	$+78.352$ &	113.2	&	0.54	&	$1050\pm 10$	&	$38.9\pm 0.8$	&	$-1.42\pm 0.08$	&	$134\pm 19$	& 9512;  9530;  9534 \\
090429B	&	210.672 &	$+32.167$ &	5.5	&	9.4	&	$3.1\pm 0.3$	&	$1.6\pm 0.2$	&	$-0.47\pm 0.77$	&	$42.1\pm 5.6$	&	 9281;  9290	\\
090424	&	189.531 &	$+16.829$ &	48	&	0.544	&	$210\pm 0$	&	$71\pm 2$	&	$-1.19\pm 0.15$	&	$108.6\pm 20.3$	&	 9223;  9231	\\
090423	&	148.891 &	$+18.165$ &	10.3	&	8	&	$5.9\pm 0.4$	&	$1.7\pm 0.2$	&	$-0.80\pm 0.50$	&	$48.6\pm 6.2$	& 9198;  9204;  9241 \\
081222	&	22.748 	&	$-34.095$ &	24	&	2.77	&	$48\pm 1$	&	$7.7\pm 0.2$	&	$-1.08\pm 0.15$	&	$131\pm 31$	&	 8691;  8709	\\
081221	&	15.801 	&	$-24.542$ &	34	&	2.26	&	$181\pm 3$	&	$18.2\pm 0.5$	&	$-1.21\pm 0.13$	&	$69.9\pm 3.9$	&	 8687;  8708	\\
081121	&	89.282 	&	$-60.612$ &	14	&	2.512	&	$41\pm 3$	&	$4.4\pm 1.0$	&	$-0.43\pm 0.54$	&	$123\pm 69$	&	 8537;  8539	\\
080916A	&	336.289 &	$-57.026$ &	60	&	0.689	&	$40\pm 1$	&	$2.7\pm 0.2$	&	$-1.17\pm 0.21$	&	$94.6\pm 23.0$	&	 8237;  8243	\\
080913	&	65.741 	&	$-25.127$ &	8	&	6.44	&	$5.6\pm 0.6$	&	$1.4\pm 0.2$	&	$-0.46\pm 0.70$	&	$93.1\pm 56.1$	&	 8217;  8222	\\
080605	&	262.130 &	$+4.010$ 	&	20	&	1.6398	&	$133\pm 2$	&	$19.9\pm 0.6$	&	$-1.11\pm 0.14$	&	$223\pm 133$	&	 7828;  7839;  7841	\\
080603B	&	176.554 &	$+68.061$ &	60	&	2.69	&	$24\pm 1$	&	$3.5\pm 0.2$	&	$-1.21\pm 0.30$	&	$71\pm 16$	&	 7794;  7806	\\
080413B	&	326.138 &	$-19.981$ &	8	&	1.1	&	$32\pm 1$	&	$18.7\pm 0.8$	&	$-1.26\pm 0.27$	&	$73.3\pm 15.8$	&	 7598;  7606;  7610	\\
080207	&	207.514 &	$+7.492$ 	&	340	&	2.0858	&	$61\pm 2$	&	$1.0\pm 0.3$	&	$-1.17\pm 0.27$	&	$107.8\pm 72.5$	&	 7264;  7272;  7277	\\
071010B	&	150.531 &	$+45.733$ &	35.7	&	0.947	&	$44\pm 1$	&	$7.7\pm 0.3$	&	$-1.53\pm 0.22$	&	$52.0\pm 6.4$	&	 6871;  6877	\\
070521	&	242.659 &	$+30.260$ &	37.9	&	0.553	&	$80.10\pm 1.77$	&	$6.53\pm 0.27$	&	$-1.10\pm 0.17$	&	$195\pm 123$	&	 6431;  6440 \\
070508	&	312.832 &	$-78.382$ &	20.9	&	0.82	&	$196.00\pm 2.73$	&	$24.10\pm 0.61$	&	$-1.14\pm 0.12$	&	$258\pm 134$	&	6383; 6390\\
060927	&	329.547 &	$+5.370$ 	&	22.5	&	5.6	&	$11.30\pm 0.68$	&	$2.70\pm 0.17$	&	$-0.93\pm 0.38$	&	$71.7\pm 17.6$	&	 5627;  5639	\\
060707	&	357.069 &	$-17.904$ &	66.2	&	3.43	&	$16.00\pm 1.51$	&	$1.01\pm 0.23$	&	$-0.66\pm 0.63$	&	$66.0\pm 17.5$	& 5285;  5289 \\
060206	&	202.933 &	$+35.046$ &	7.6	&	4.045	&	$8.31\pm 0.42$	&	$2.79\pm 0.17$	&	$-1.06\pm 0.34$	&	$75.4\pm 19.5$	&	 4682;  4697	\\
060115	&	54.007 	&	$+17.339$ &	139.6	&	3.53	&	$17.1\pm 1.5$	&	$0.87\pm 0.12$	&	$-1.00\pm 0.50$	&	$62.0\pm 20.5$	&	 4509;  4518	\\
050525A	&	278.142 &	$+26.335$ &	8.8	&	0.606	&	$153.00\pm 2.21$	&	$41.70\pm 0.94$	&	$-1.00\pm 0.10$	&	$79\pm 4$	& 3466;  3467;  3479 \\
\hline
\end{tabular}
\end{table*}
The errors in this table are of 90\% confidence level (C.L.). All of the 44 GRBs are long GRBs with duration $T_{90}>2$ seconds\footnote{The duration of a GRB usually characterized by $T_{90}$, in which time from 5\% to 95\% photons are recorded. The duration of GRB 130925A is not measured, but it is very likely to be a long GRB, because all the other GRBs having reliable CPL fit are long GRBs, and GRB 130925A follows the Amati relation and Yonetoku relation of the remaining GRBs (see section \ref{sec:correlation}).}. The observed photon fluence spans three orders of magnitude, while the peak photon flux spans two orders of magnitude. Figure \ref{fig:histogram}(a) is the histogram of redshift of the 44 GRBs.
The redshift of our sample ranges from 0.347 to 9.4, with a median value 1.9. Figure \ref{fig:histogram}(b) is the histogram of the low-energy power-law index. A Gauss fit to the histogram gives the average value $\bar{\alpha}=-1.05\pm 0.14$ and standard deviation $\sigma_{\alpha}=0.38\pm 0.14$ (95\% C.L.). Two GRBs (GRB 131117A and GRB 130612A) have positive index. Figure \ref{fig:histogram}(c) is the distribution of the observed peak energy $E_p$, ranging from 19.2 keV to 258 keV, with a median value 71.4 keV. Only 9\% (4 out of 44) GRBs have peak energy above the BAT energy band. All the remaining GRBs have peak energy well within the BAT energy range. This is one reason why the high-energy power-law index can't be well constrained. In Figure \ref{fig:histogram}(d) we plot the histogram of the rest frame peak energy $E_{p,i}=E_p\times (1+z)$, which ranges from 38 keV to 693 keV. The distribution of $E_{p,i}$ is more homogeneous compared to that of $E_p$\,.

\section{The luminosity correlations}\label{sec:correlation}


The most investigated GRB luminosity correlations are those between the total radiated energy or luminosity (calculated in a fixed energy band) and the spectral peak energy. Here we focus on discussing two luminosity correlations, i.e., the Amati relation ($E_p-E_{\rm iso}$) and Yonetoku relation ($E_p-L$).

The Amati relation is a linear (in the logarithmic scale) correlation between isotropic equivalent energy and peak energy of $\nu F_{\nu}$ spectrum in the comoving frame \citep{Amati:2002,Amati:2003,Amati:2006}. It was first discovered by \citet{Amati:2002} in 12 long BeppoSAX GRBs and was confirmed in larger samples \citep{Amati:2003,Amati:2006}. Later it was found that short GRBs do not follow the same Amati relation of long GRBs \citep{Amati:2009,Amati:2010}. The physical origin of Amati relation has been discussed but selection effect couldn't be excluded. Amati relation is one of the most used relations in calibrating the distance of GRBs, although there are still some controversies on its validity \citep{Li:2007,Dainotti:2013,Lin:2015a}. We can parameterize Amati relation as
\begin{equation}\label{eq:amati}
  \log\frac{E_{\rm iso}}{{\rm erg}}=a+b\log\frac{E_{p,i}}{300~{\rm keV}},
\end{equation}
where $E_{p,i}=E_p\times (1+z)$ is the peak energy in the rest frame, and
\begin{equation}\label{eq:iso_energy}
  E_{\rm iso}=4\pi d_L^2S_{\rm bolo}(1+z)^{-1}
\end{equation}
is the isotropic equivalent energy in the rest frame $1-10,000$ keV energy band, and $d_L$ is the luminosity distance. In the concordance cosmological model, the luminosity distance is given by
\begin{equation}\label{lumi_distance}
  d_L(z)=(1+z)\frac{c}{H_0}\int_0^{z} \frac{dz}{\sqrt{\Omega_M(1+z)^3+(1-\Omega_M)}}.
\end{equation}
The bolometric fluence $S_{\rm bolo}$ is calculated from the observed photon fluence $S$ and GRB spectra $N(E)$ in the rest frame $1-10,000$ keV energy band \citep{Schaefer:2007},
\begin{equation}
  S_{\rm bolo}=S\times \frac{\int_{1/(1+z)}^{10^4/(1+z)}EN(E)dE}{\int_{E_{\rm min}}^{E_{\rm max}}EN(E)dE},
\end{equation}
where $E_{\rm min}=15$ keV and $E_{\rm max}=150$ keV are the lower and upper boundary of the BAT energy band, respectively. The bolometric fluence $S_{\rm bolo}$ , as well as the isotropic equivalent energy $E_{\rm iso}$, depend on the GRB spectra. We calculate $S_{\rm bolo}$ and $E_{\rm iso}$ in both cases that GRB spectra are modeled by Band function and CPL. Since the high-energy power-law index of the Band function can't be measured, we fix it to be $-2.2$, the peak of BATSE catalog \citep{Preece:2000}. The final results are listed in columns $(2)-(5)$ in Table \ref{tab:parameters2}.
\begin{table*}
\centering
\caption{\small{The bolometric fluence, isotropic equivalent energy, and distance modulus calibrated using the Amati relation. Column (1): the GRB identifier. Columns (2) and (3): the bolometric fluence in unit of $10^{-7}$ erg/cm$^2$. Columns (4) and (5): the isotropic equivalent energy in unit of $10^{52}$ ergs. Columns (6) and (7): the distance modulus calibrated using the Amati relation. The superscripts ``\,CPL" and ``\,Band" mean that the quantities are calculated using the CPL and Band function, respectively. All the errors in this table are of $1\sigma$ C.L.}}\label{tab:parameters2}
\begin{tabular}{lllllll}
\hline
(1)	&	(2)	&	(3)	&	(4)	&	(5)	&	(6)	&	(7)	\\
GRBs	&$S_{\rm bolo}^{\rm CPL}$	& $S_{\rm bolo}^{\rm Band}$	& $E_{\rm iso}^{\rm CPL}$	& $E_{\rm iso}^{\rm Band}$	& $\mu^{\rm CPL}$	& $\mu^{\rm Band}$\\
\hline
141220A	& $39.83 \pm 0.93 $	& $70.20 \pm 1.65 $	& $1.83 \pm 0.04 $	& $3.24 \pm 0.08 $	& $46.023 \pm 1.245 $	& $45.864 \pm 1.200 $	\\
140518A	& $14.29 \pm 0.87 $	& $21.08 \pm 1.29 $	& $5.93 \pm 0.36 $	& $8.74 \pm 0.53 $	& $48.019 \pm 1.225 $	& $48.053 \pm 1.179 $	\\
140515A	& $8.43 \pm 0.52 $	& $12.16 \pm 0.75 $	& $5.42 \pm 0.34 $	& $7.85 \pm 0.49 $	& $49.316 \pm 1.236 $	& $49.381 \pm 1.190 $	\\
140206A	& $254.35 \pm 2.91 $	& $385.53 \pm 4.41 $	& $43.78 \pm 0.50 $	& $66.34 \pm 0.76 $	& $44.883 \pm 1.223 $	& $44.899 \pm 1.176 $	\\
131117A	& $2.85 \pm 0.28 $	& $4.86 \pm 0.47 $	& $1.00 \pm 0.10 $	& $1.68 \pm 0.16 $	& $49.541 \pm 1.229 $	& $49.433 \pm 1.183 $	\\
130925A	& $987.08 \pm 14.68 $	& $1279.91 \pm 19.04 $	& $3.00 \pm 0.04 $	& $3.89 \pm 0.06 $	& $39.950 \pm 1.389 $	& $40.075 \pm 1.343 $	\\
130701A	& $64.35 \pm 0.89 $	& $109.31 \pm 1.51 $	& $2.29 \pm 0.03 $	& $3.89 \pm 0.05 $	& $45.034 \pm 1.227 $	& $44.908 \pm 1.180 $	\\
130612A	& $2.84 \pm 0.38 $	& $4.75 \pm 0.63 $	& $0.28 \pm 0.04 $	& $0.48 \pm 0.06 $	& $48.175 \pm 1.266 $	& $48.034 \pm 1.218 $	\\
130420A	& $139.49 \pm 3.59 $	& $192.50 \pm 4.96 $	& $6.22 \pm 0.16 $	& $8.59 \pm 0.22 $	& $43.239 \pm 1.284 $	& $43.312 \pm 1.235 $	\\
120923A	& $3.95 \pm 0.60 $	& $5.88 \pm 0.90 $	& $3.85 \pm 0.59 $	& $5.80 \pm 0.88 $	& $50.561 \pm 1.243 $	& $50.584 \pm 1.196 $	\\
120922A	& $127.40 \pm 8.77 $	& $166.51 \pm 11.46 $	& $27.13 \pm 1.87 $	& $35.46 \pm 2.44 $	& $44.765 \pm 1.233 $	& $44.916 \pm 1.185 $	\\
120811C	& $53.33 \pm 3.25 $	& $74.43 \pm 4.54 $	& $8.84 \pm 0.54 $	& $12.34 \pm 0.75 $	& $45.597 \pm 1.235 $	& $45.677 \pm 1.188 $	\\
120802A	& $29.81 \pm 2.87 $	& $43.13 \pm 4.15 $	& $8.82 \pm 0.85 $	& $12.77 \pm 1.23 $	& $47.134 \pm 1.243 $	& $47.191 \pm 1.198 $	\\
120724A	& $10.47 \pm 1.03 $	& $15.23 \pm 1.50 $	& $0.60 \pm 0.06 $	& $0.87 \pm 0.09 $	& $46.011 \pm 1.303 $	& $46.026 \pm 1.255 $	\\
120326A	& $46.43 \pm 3.27 $	& $66.04 \pm 4.65 $	& $3.82 \pm 0.27 $	& $5.44 \pm 0.38 $	& $45.104 \pm 1.254 $	& $45.155 \pm 1.206 $	\\
110726A	& $2.87 \pm 0.24 $	& $4.97 \pm 0.41 $	& $0.08 \pm 0.01 $	& $0.14 \pm 0.01 $	& $47.554 \pm 1.274 $	& $47.397 \pm 1.226 $	\\
110715A	& $213.84 \pm 2.21 $	& $322.87 \pm 3.34 $	& $3.85 \pm 0.04 $	& $5.81 \pm 0.06 $	& $43.689 \pm 1.226 $	& $43.694 \pm 1.179 $	\\
110503A	& $170.72 \pm 4.16 $	& $273.55 \pm 6.67 $	& $11.51 \pm 0.28 $	& $18.44 \pm 0.45 $	& $44.842 \pm 1.249 $	& $44.794 \pm 1.204 $	\\
110422A	& $743.18 \pm 11.05 $	& $1169.78 \pm 17.40 $	& $59.50 \pm 0.88 $	& $93.65 \pm 1.39 $	& $43.501 \pm 1.223 $	& $43.479 \pm 1.176 $	\\
100816A	& $38.98 \pm 1.19 $	& $65.91 \pm 2.01 $	& $0.67 \pm 0.02 $	& $1.14 \pm 0.03 $	& $45.907 \pm 1.258 $	& $45.799 \pm 1.214 $	\\
091029	& $43.84 \pm 1.11 $	& $60.14 \pm 1.53 $	& $7.64 \pm 0.19 $	& $10.49 \pm 0.27 $	& $46.255 \pm 1.235 $	& $46.364 \pm 1.189 $	\\
091018	& $35.10 \pm 1.53 $	& $44.45 \pm 1.94 $	& $0.89 \pm 0.04 $	& $1.12 \pm 0.05 $	& $43.794 \pm 1.443 $	& $43.940 \pm 1.399 $	\\
090926B	& $94.19 \pm 1.57 $	& $174.84 \pm 2.92 $	& $3.85 \pm 0.06 $	& $7.15 \pm 0.12 $	& $44.562 \pm 1.228 $	& $44.336 \pm 1.180 $	\\
090618	& $2122.01 \pm 12.32 $	& $3024.54 \pm 17.56 $	& $16.19 \pm 0.09 $	& $23.07 \pm 0.13 $	& $40.953 \pm 1.225 $	& $41.019 \pm 1.178 $	\\
090429B	& $4.00 \pm 0.24 $	& $5.71 \pm 0.34 $	& $4.54 \pm 0.27 $	& $6.48 \pm 0.38 $	& $50.673 \pm 1.227 $	& $50.757 \pm 1.180 $	\\
090424	& $358.90 \pm 0.00 $	& $568.15 \pm 0.00 $	& $2.78 \pm 0.00 $	& $4.40 \pm 0.00 $	& $42.655 \pm 1.234 $	& $42.601 \pm 1.187 $	\\
090423	& $8.00 \pm 0.33 $	& $11.50 \pm 0.48 $	& $7.25 \pm 0.30 $	& $10.42 \pm 0.43 $	& $49.762 \pm 1.225 $	& $49.839 \pm 1.178 $	\\
081222	& $85.28 \pm 1.08 $	& $123.41 \pm 1.57 $	& $15.04 \pm 0.19 $	& $21.77 \pm 0.28 $	& $46.382 \pm 1.235 $	& $46.456 \pm 1.189 $	\\
081221	& $285.91 \pm 2.89 $	& $430.14 \pm 4.35 $	& $35.50 \pm 0.36 $	& $53.41 \pm 0.54 $	& $44.054 \pm 1.220 $	& $44.063 \pm 1.173 $	\\
081121	& $62.92 \pm 2.81 $	& $108.59 \pm 4.84 $	& $9.39 \pm 0.42 $	& $16.21 \pm 0.72 $	& $46.488 \pm 1.280 $	& $46.366 \pm 1.238 $	\\
080916A	& $65.19 \pm 0.99 $	& $104.26 \pm 1.59 $	& $0.82 \pm 0.01 $	& $1.32 \pm 0.02 $	& $44.551 \pm 1.241 $	& $44.485 \pm 1.194 $	\\
080913	& $7.53 \pm 0.49 $	& $11.88 \pm 0.78 $	& $4.98 \pm 0.33 $	& $7.88 \pm 0.51 $	& $50.135 \pm 1.307 $	& $50.120 \pm 1.264 $	\\
080605	& $313.06 \pm 2.87 $	& $438.19 \pm 4.02 $	& $21.77 \pm 0.20 $	& $30.46 \pm 0.28 $	& $44.778 \pm 1.296 $	& $44.893 \pm 1.254 $	\\
080603B	& $38.02 \pm 0.97 $	& $56.41 \pm 1.43 $	& $6.38 \pm 0.16 $	& $9.47 \pm 0.24 $	& $46.534 \pm 1.227 $	& $46.562 \pm 1.181 $	\\
080413B	& $51.85 \pm 0.99 $	& $80.08 \pm 1.53 $	& $1.68 \pm 0.03 $	& $2.59 \pm 0.05 $	& $44.996 \pm 1.240 $	& $44.965 \pm 1.193 $	\\
080207	& $103.81 \pm 2.08 $	& $153.66 \pm 3.07 $	& $11.18 \pm 0.22 $	& $16.55 \pm 0.33 $	& $45.514 \pm 1.300 $	& $45.551 \pm 1.260 $	\\
071010B	& $82.49 \pm 1.14 $	& $117.70 \pm 1.63 $	& $1.98 \pm 0.03 $	& $2.83 \pm 0.04 $	& $43.944 \pm 1.257 $	& $43.989 \pm 1.209 $	\\
070521	& $173.90 \pm 2.34 $	& $266.09 \pm 3.59 $	& $1.39 \pm 0.02 $	& $2.13 \pm 0.03 $	& $44.104 \pm 1.290 $	& $44.103 \pm 1.249 $	\\
070508	& $503.41 \pm 4.28 $	& $720.52 \pm 6.12 $	& $9.06 \pm 0.08 $	& $12.96 \pm 0.11 $	& $43.608 \pm 1.273 $	& $43.692 \pm 1.230 $	\\
060927	& $15.99 \pm 0.59 $	& $23.77 \pm 0.87 $	& $8.64 \pm 0.32 $	& $12.83 \pm 0.47 $	& $48.760 \pm 1.236 $	& $48.804 \pm 1.189 $	\\
060707	& $20.73 \pm 1.19 $	& $34.40 \pm 1.98 $	& $5.21 \pm 0.30 $	& $8.65 \pm 0.50 $	& $47.514 \pm 1.232 $	& $47.422 \pm 1.186 $	\\
060206	& $12.42 \pm 0.38 $	& $18.49 \pm 0.57 $	& $4.06 \pm 0.13 $	& $6.06 \pm 0.19 $	& $48.503 \pm 1.232 $	& $48.537 \pm 1.186 $	\\
060115	& $24.48 \pm 1.31 $	& $37.63 \pm 2.01 $	& $6.46 \pm 0.34 $	& $9.91 \pm 0.53 $	& $47.310 \pm 1.239 $	& $47.304 \pm 1.194 $	\\
050525A	& $224.12 \pm 1.97 $	& $383.56 \pm 3.38 $	& $2.17 \pm 0.02 $	& $3.71 \pm 0.03 $	& $42.900 \pm 1.241 $	& $42.753 \pm 1.193 $	\\
\hline
\end{tabular}
\end{table*}
Errors are given in $1\sigma$ C.L. We only consider the error propagation from $S$. As is expected, the isotropic equivalent energy calculated from Band function is much larger than that from CPL, since CPL falls much faster than Band function above the breaking energy.


The Yonetoku relation is first found by \citet{Yonetoku:2004} from the combination of 12 BeppoSAX GRBs and 11 BATSE GRBs. It is similar to the Amati relation except that the isotropic equivalent energy $E_{\rm iso}$ is replaced by the isotropic peak luminosity $L$, i.e.,
\begin{equation}
  \log\frac{L}{{\rm erg~s}^{-1}}=a+b\log\frac{E_{p,i}}{300~{\rm keV}}.
\end{equation}
The isotropic peak luminosity $L$ can be calculated from the bolometric peak flux $P_{\rm bolo}$ as
\begin{equation}\label{eq:iso_luminosity}
  L=4\pi d_L^2P_{\rm bolo},
\end{equation}
while the bolometric peak flux $P_{\rm bolo}$ is calculated from the observed peak photon flux $P$ in the rest frame $1-10,000$ keV energy band by integrating over the GRB spectra  \citep{Schaefer:2007},
\begin{equation}
  P_{\rm bolo}=P\times \frac{\int_{1/(1+z)}^{10^4/(1+z)}EN(E)dE}{\int_{E_{\rm min}}^{E_{\rm max}}N(E)dE}.
\end{equation}
We also calculate $P_{\rm bolo}$ and $L$ in both the Band and CPL spectra cases, and list the results in columns $(2)-(5)$ in Table \ref{tab:parameters3}.
\begin{table*}
\centering
\caption{\small{The bolometric peak flux, isotropic peak luminosity, and distance modulus calibrated using the Yonetoku relation. Column (1): the GRB identifier. Columns (2) and (3): the bolometric peak flux in unit of $10^{-7}$ erg/cm$^2$/s. Columns (4) and (5): the isotropic peak luminosity in unit of $10^{51}$ erg/s. Columns (6) and (7): the distance modulus calibrated using the Yonetoku relation. The superscripts ``\,CPL" and ``\,Band" mean that the quantities are calculated using the CPL and Band function, respectively. All the errors in this table are of $1\sigma$ C.L.}}\label{tab:parameters3}
\begin{tabular}{lllllll}
\hline
(1)	&	(2)	&	(3)	&	(4)	&	(5)	&	(6)	&	(7)	\\
GRBs	&$P_{\rm bolo}^{\rm CPL}$	& $P_{\rm bolo}^{\rm Band}$	& $L^{\rm CPL}$	& $L^{\rm Band}$	& $\mu^{\rm CPL}$	& $\mu^{\rm Band}$\\
\hline
141220A	& $11.39 \pm 0.55 $	& $20.08 \pm 0.96 $	& $12.18 \pm 0.58 $	& $21.47 \pm 1.03 $	& $45.297 \pm 1.249 $	& $45.172 \pm 1.214 $	\\
140518A	& $0.85 \pm 0.05 $	& $1.34 \pm 0.08 $	& $20.02 \pm 1.22 $	& $31.76 \pm 1.94 $	& $47.971 \pm 1.193 $	& $47.960 \pm 1.156 $	\\
140515A	& $0.80 \pm 0.05 $	& $1.21 \pm 0.08 $	& $37.96 \pm 2.57 $	& $57.26 \pm 3.88 $	& $48.751 \pm 1.220 $	& $48.796 \pm 1.184 $	\\
140206A	& $22.93 \pm 0.36 $	& $34.78 \pm 0.55 $	& $147.17 \pm 2.31 $	& $223.21 \pm 3.51 $	& $45.117 \pm 1.187 $	& $45.156 \pm 1.150 $	\\
131117A	& $0.47 \pm 0.04 $	& $0.93 \pm 0.08 $	& $8.38 \pm 0.73 $	& $16.64 \pm 1.45 $	& $48.442 \pm 1.195 $	& $48.187 \pm 1.159 $	\\
130925A	& $10.49 \pm 0.53 $	& $13.67 \pm 0.69 $	& $0.43 \pm 0.02 $	& $0.56 \pm 0.03 $	& $42.162 \pm 1.448 $	& $42.359 \pm 1.412 $	\\
130701A	& $18.53 \pm 0.46 $	& $31.58 \pm 0.79 $	& $14.22 \pm 0.35 $	& $24.24 \pm 0.61 $	& $44.145 \pm 1.192 $	& $44.055 \pm 1.155 $	\\
130612A	& $1.11 \pm 0.12 $	& $2.24 \pm 0.24 $	& $3.34 \pm 0.36 $	& $6.77 \pm 0.73 $	& $46.195 \pm 1.235 $	& $45.917 \pm 1.198 $	\\
130420A	& $3.78 \pm 0.14 $	& $5.51 \pm 0.20 $	& $3.88 \pm 0.14 $	& $5.65 \pm 0.20 $	& $44.214 \pm 1.257 $	& $44.293 \pm 1.219 $	\\
120923A	& $0.44 \pm 0.04 $	& $0.73 \pm 0.07 $	& $40.96 \pm 4.16 $	& $68.35 \pm 6.94 $	& $49.621 \pm 1.212 $	& $49.556 \pm 1.176 $	\\
120922A	& $2.41 \pm 0.15 $	& $3.25 \pm 0.20 $	& $21.08 \pm 1.29 $	& $28.40 \pm 1.73 $	& $45.991 \pm 1.191 $	& $46.156 \pm 1.153 $	\\
120811C	& $4.34 \pm 0.13 $	& $6.28 \pm 0.19 $	& $26.41 \pm 0.79 $	& $38.25 \pm 1.14 $	& $45.364 \pm 1.198 $	& $45.451 \pm 1.161 $	\\
120802A	& $3.01 \pm 0.12 $	& $4.45 \pm 0.18 $	& $42.77 \pm 1.74 $	& $63.22 \pm 2.57 $	& $46.755 \pm 1.233 $	& $46.821 \pm 1.198 $	\\
120724A	& $0.44 \pm 0.09 $	& $0.80 \pm 0.16 $	& $0.63 \pm 0.13 $	& $1.14 \pm 0.23 $	& $46.356 \pm 1.300 $	& $46.197 \pm 1.262 $	\\
120326A	& $4.84 \pm 0.13 $	& $7.18 \pm 0.19 $	& $11.17 \pm 0.30 $	& $16.54 \pm 0.44 $	& $44.682 \pm 1.220 $	& $44.743 \pm 1.182 $	\\
110726A	& $0.79 \pm 0.10 $	& $1.48 \pm 0.18 $	& $0.46 \pm 0.06 $	& $0.87 \pm 0.11 $	& $46.304 \pm 1.257 $	& $46.108 \pm 1.220 $	\\
110715A	& $71.55 \pm 0.89 $	& $108.02 \pm 1.34 $	& $23.43 \pm 0.29 $	& $35.37 \pm 0.44 $	& $42.907 \pm 1.194 $	& $42.949 \pm 1.157 $	\\
110503A	& $1.87 \pm 0.05 $	& $3.00 \pm 0.08 $	& $3.29 \pm 0.09 $	& $5.28 \pm 0.14 $	& $47.696 \pm 1.257 $	& $47.675 \pm 1.222 $	\\
110422A	& $46.31 \pm 0.92 $	& $72.89 \pm 1.45 $	& $102.69 \pm 2.04 $	& $161.63 \pm 3.21 $	& $44.524 \pm 1.186 $	& $44.523 \pm 1.149 $	\\
100816A	& $18.69 \pm 0.42 $	& $31.60 \pm 0.71 $	& $5.82 \pm 0.13 $	& $9.84 \pm 0.22 $	& $44.979 \pm 1.281 $	& $44.899 \pm 1.247 $	\\
091029	& $2.10 \pm 0.07 $	& $2.91 \pm 0.10 $	& $13.76 \pm 0.47 $	& $19.03 \pm 0.64 $	& $46.832 \pm 1.218 $	& $46.970 \pm 1.182 $	\\
091018	& $13.98 \pm 0.33 $	& $18.89 \pm 0.45 $	& $6.96 \pm 0.16 $	& $9.41 \pm 0.22 $	& $41.541 \pm 1.549 $	& $41.698 \pm 1.515 $	\\
090926B	& $3.10 \pm 0.18 $	& $5.83 \pm 0.33 $	& $2.84 \pm 0.16 $	& $5.34 \pm 0.31 $	& $45.923 \pm 1.190 $	& $45.725 \pm 1.153 $	\\
090618	& $55.98 \pm 0.70 $	& $79.79 \pm 1.00 $	& $6.58 \pm 0.08 $	& $9.37 \pm 0.12 $	& $43.072 \pm 1.190 $	& $43.176 \pm 1.153 $	\\
090429B	& $1.19 \pm 0.09 $	& $1.90 \pm 0.15 $	& $140.28 \pm 10.69 $	& $224.69 \pm 17.13 $	& $48.601 \pm 1.191 $	& $48.581 \pm 1.155 $	\\
090424	& $88.74 \pm 1.52 $	& $140.49 \pm 2.41 $	& $10.61 \pm 0.18 $	& $16.80 \pm 0.29 $	& $42.200 \pm 1.204 $	& $42.190 \pm 1.167 $	\\
090423	& $1.42 \pm 0.10 $	& $2.17 \pm 0.16 $	& $115.40 \pm 8.28 $	& $176.88 \pm 12.69 $	& $48.410 \pm 1.190 $	& $48.438 \pm 1.153 $	\\
081222	& $10.56 \pm 0.17 $	& $15.28 \pm 0.24 $	& $70.22 \pm 1.11 $	& $101.61 \pm 1.61 $	& $46.446 \pm 1.211 $	& $46.536 \pm 1.175 $	\\
081221	& $19.28 \pm 0.32 $	& $29.27 \pm 0.49 $	& $78.06 \pm 1.31 $	& $118.47 \pm 1.98 $	& $44.406 \pm 1.179 $	& $44.443 \pm 1.142 $	\\
081121	& $5.93 \pm 0.82 $	& $10.24 \pm 1.42 $	& $31.10 \pm 4.31 $	& $53.67 \pm 7.44 $	& $46.832 \pm 1.337 $	& $46.731 \pm 1.305 $	\\
080916A	& $3.16 \pm 0.14 $	& $5.05 \pm 0.23 $	& $0.67 \pm 0.03 $	& $1.08 \pm 0.05 $	& $45.736 \pm 1.218 $	& $45.714 \pm 1.182 $	\\
080913	& $1.51 \pm 0.13 $	& $2.40 \pm 0.21 $	& $74.69 \pm 6.51 $	& $118.39 \pm 10.31 $	& $49.161 \pm 1.370 $	& $49.153 \pm 1.338 $	\\
080605	& $38.23 \pm 0.70 $	& $53.51 \pm 0.98 $	& $70.16 \pm 1.29 $	& $98.21 \pm 1.81 $	& $45.363 \pm 1.357 $	& $45.490 \pm 1.324 $	\\
080603B	& $3.73 \pm 0.13 $	& $5.58 \pm 0.19 $	& $23.11 \pm 0.81 $	& $34.56 \pm 1.20 $	& $46.440 \pm 1.202 $	& $46.493 \pm 1.165 $	\\
080413B	& $20.41 \pm 0.53 $	& $31.71 \pm 0.83 $	& $13.87 \pm 0.36 $	& $21.55 \pm 0.56 $	& $43.642 \pm 1.213 $	& $43.652 \pm 1.176 $	\\
080207	& $1.25 \pm 0.23 $	& $1.85 \pm 0.34 $	& $4.15 \pm 0.76 $	& $6.14 \pm 1.12 $	& $48.056 \pm 1.400 $	& $48.121 \pm 1.370 $	\\
071010B	& $8.95 \pm 0.21 $	& $12.93 \pm 0.31 $	& $4.19 \pm 0.10 $	& $6.05 \pm 0.14 $	& $43.787 \pm 1.222 $	& $43.875 \pm 1.183 $	\\
070521	& $11.45 \pm 0.29 $	& $17.53 \pm 0.44 $	& $1.43 \pm 0.04 $	& $2.18 \pm 0.06 $	& $45.481 \pm 1.362 $	& $45.510 \pm 1.331 $	\\
070508	& $50.66 \pm 0.78 $	& $72.50 \pm 1.12 $	& $16.59 \pm 0.26 $	& $23.74 \pm 0.37 $	& $44.653 \pm 1.311 $	& $44.755 \pm 1.278 $	\\
060927	& $2.66 \pm 0.10 $	& $4.00 \pm 0.15 $	& $94.65 \pm 3.63 $	& $142.63 \pm 5.48 $	& $47.867 \pm 1.213 $	& $47.914 \pm 1.176 $	\\
060707	& $0.91 \pm 0.13 $	& $1.55 \pm 0.22 $	& $10.12 \pm 1.41 $	& $17.27 \pm 2.40 $	& $48.170 \pm 1.219 $	& $48.080 \pm 1.184 $	\\
060206	& $2.90 \pm 0.11 $	& $4.35 \pm 0.16 $	& $47.87 \pm 1.78 $	& $71.83 \pm 2.67 $	& $47.382 \pm 1.211 $	& $47.433 \pm 1.175 $	\\
060115	& $0.83 \pm 0.07 $	& $1.30 \pm 0.11 $	& $9.85 \pm 0.83 $	& $15.49 \pm 1.30 $	& $48.203 \pm 1.233 $	& $48.201 \pm 1.197 $	\\
050525A	& $43.25 \pm 0.59 $	& $74.53 \pm 1.02 $	& $6.73 \pm 0.09 $	& $11.59 \pm 0.16 $	& $42.481 \pm 1.201 $	& $42.378 \pm 1.163 $	\\
\hline
\end{tabular}
\end{table*}
Errors are given in $1\sigma$ C.L. The errors of $P_{\rm bolo}$ and $L$ are propagated from the errors of $P$. Similarly, we can see that the peak luminosity calculated from Band function is much larger than that from CPL.

\begin{figure}
\centering
 \includegraphics[width=0.5\textwidth,height=12cm]{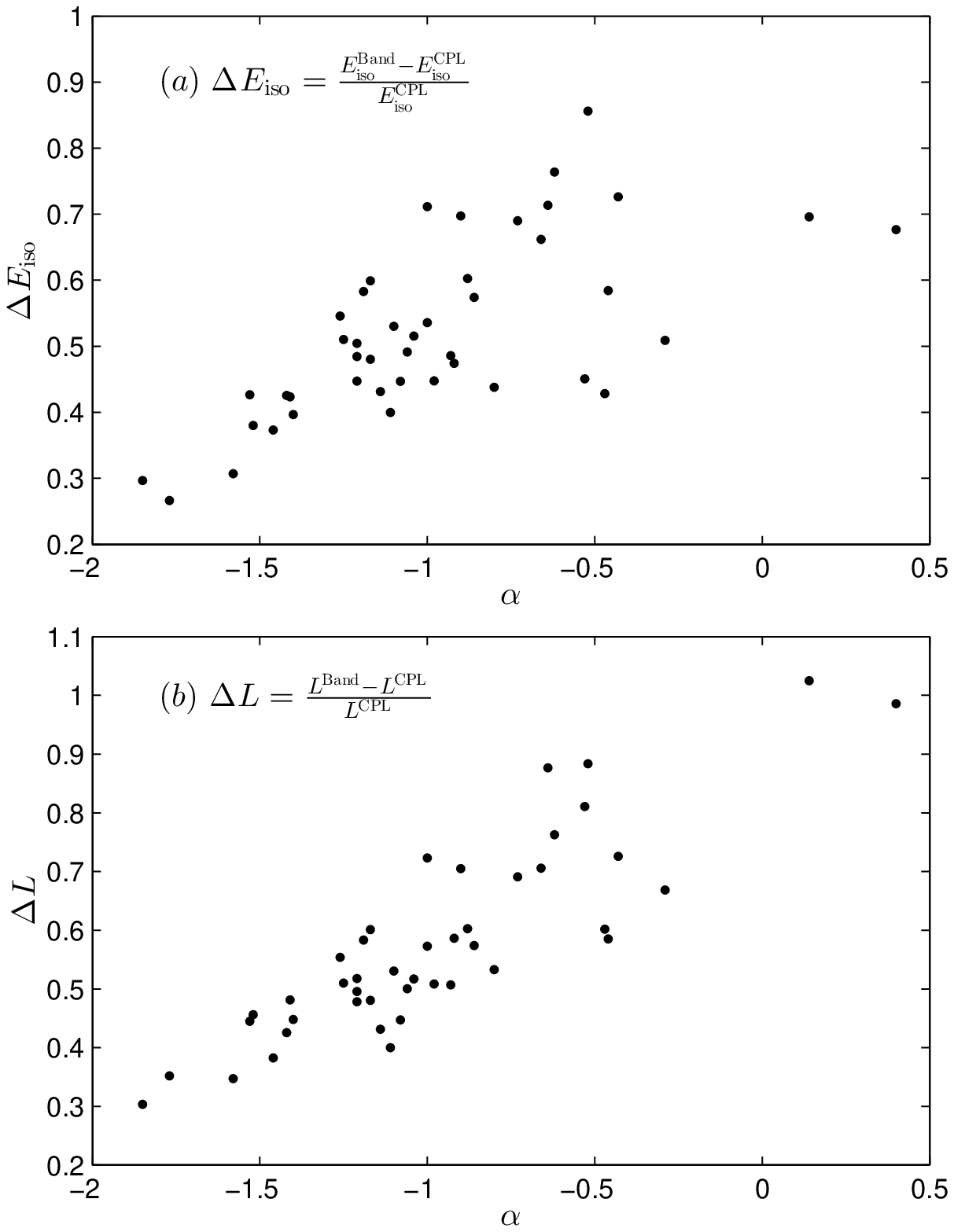}
 \caption{\small{Top panel: the correlation between $\Delta E_{\rm iso}$ and $\alpha$. Bottle panel: the correlation between $\Delta L$ and $\alpha$.}} \label{fig:alpha_dEiso}
\end{figure}

In order to see more clearly the effect of GRB spectra on the final isotropic energy (and peak luminosity), we calculate the difference of isotropic energy (and peak luminosity) derived from Band and CPL spectra,
\begin{equation}
  \begin{dcases}
    \Delta E_{\rm iso} \equiv\frac{E_{\rm iso}^{\rm Band}-E_{\rm iso}^{\rm CPL}}{E_{\rm iso}^{\rm CPL}},\\
    \Delta L \equiv\frac{L^{\rm Band}-L^{\rm CPL}}{L^{\rm CPL}}.
  \end{dcases}
\end{equation}
We plot $\Delta E_{\rm iso}$ and $\Delta L$ versus the power-law index $\alpha$ in Figure \ref{fig:alpha_dEiso}.
We can see that both $\Delta E_{\rm iso}$ and $\Delta L$ are positively correlated with $\alpha$. When $\alpha$ is close to $-2$, both $\Delta E_{\rm iso}$ and $\Delta L$ are small. This is because the difference between Band function and CPL model becomes smaller when $\alpha$ approaches $\beta$.


Both Amati relation and Yonetoku relation have the linear form
\begin{equation}
  y=a+bx.
\end{equation}
The best-fitting parameters ($a,b$) can be derived using the likelihood method presented in \citet{DAgostini:2005}, i.e., by maximizing the following likelihood:
\begin{eqnarray}\label{eq:likelihood}\nonumber
  \mathcal{L}(\sigma_{\rm int},a,b) &\propto& \prod_i\frac{1}{\sqrt{\sigma_{\rm int}^2+\sigma_{y_i}^2+b^2\sigma_{x_i}^2}}\\
  &\times& \exp\left[-\frac{(y_i-a-bx_i)^2}{2(\sigma_{\rm int}^2+\sigma_{y_i}^2+b^2\sigma_{x_i}^2)}\right],
\end{eqnarray}
where $\sigma_{\rm int}$ is the intrinsic scatter, which represents the unknown errors. The introduction of intrinsic scatter is necessary since it dominates over the measurement error.

\begin{table*}
\centering
\caption{\small{The best-fitting parameters of $E_p-E_{\rm iso}$ relation and $E_p-L$ relation. The errors are of $1\sigma$ C.L. The last column gives the Pearson's linear correlation coefficient.}}\label{tab:parameters4}
\begin{tabular}{lcccc}
\hline
 & $\sigma_{\rm int}$ & $a$ & $b$ & $\rho$ \\
\hline
$E_p-E_{\rm iso}$ (CPL) & $0.480\pm 0.052$ & $52.764\pm 0.079$ & $1.021\pm 0.258$ & 0.516\\
$E_p-E_{\rm iso}$ (Band) & $0.462\pm 0.050$ & $52.949\pm 0.076$ & $1.047\pm 0.248$ & 0.541\\
\hline
$E_p-L$ (CPL) & $0.463\pm 0.052$ & $52.323\pm 0.078$ & $1.650\pm 0.259$ & 0.691\\
$E_p-L$ (Band) & $0.449\pm 0.050$ & $52.519\pm 0.076$ & $1.652\pm 0.251$ & 0.702\\
\hline
\end{tabular}
\end{table*}

\begin{figure*}
  \centering
 \includegraphics[width=0.95\textwidth]{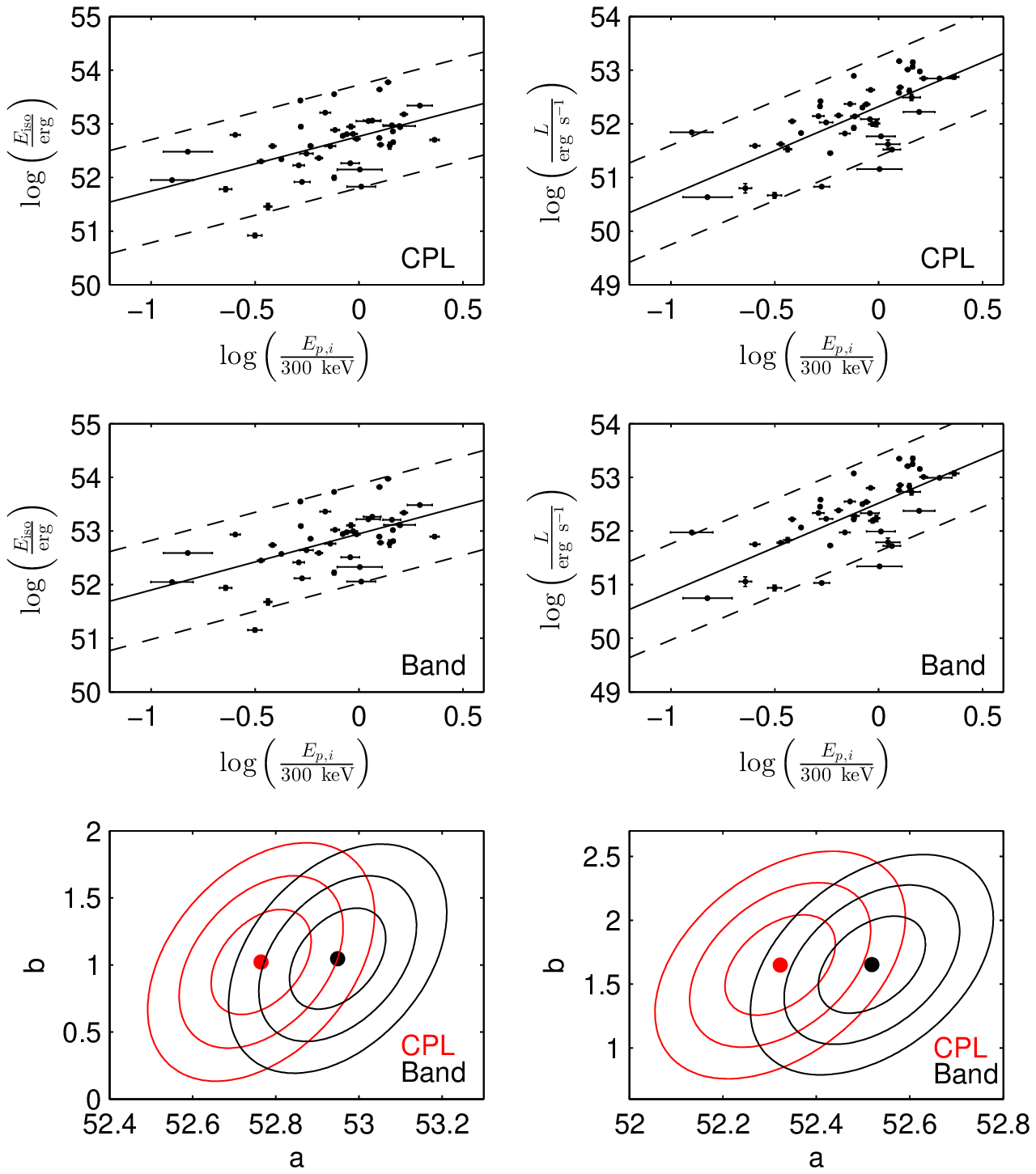}
 \caption{\small{The $E_p-E_{\rm iso}$ relation (left three panels) and $E_p-L$ relation (right three panels). The solid lines are the best-fitting results, and the dotted lines represent the $2\sigma$ scatters. Top two panels: in the case of CPL spectrum. Middle two panels: in the case of Band spectrum. Bottle two panels: the $1\sigma$, $2\sigma$ and $3\sigma$ contours in the ($a,b$) plane, red in the CPL case and black in the Band case.}}\label{fig:correlations}
\end{figure*}

We plot the Amati relation and Yonetoku relation in Figure \ref{fig:correlations}, and list the best-fitting parameters in Table \ref{tab:parameters4}. From Table \ref{tab:parameters4}, we can see that the intercept parameters of both $E_p-E_{\rm iso}$ relation and $E_p-L$ relation calculated from the Band spectrum are moderately larger than that from CPL spectrum, while the slope parameters show no significant difference. This is consistent with the results of \citet{Cabrera:2007}\footnote{In the work of \citet{Cabrera:2007}, the authors parameterized the Amati relation as $\log(E_{p,i}/{\rm keV})=a'+b'\log(E_{\rm iso}/{\rm erg})$. They found that $a'$ in the Band case is smaller than in the CPL cases. A smaller $a'$ corresponds to a larger $a$ in our parametrization.}. The best-fitting parameters of Yonetoku relation are well consistent with the results of \citet{Schaefer:2007}. However, the slope parameters of Amati relation are somehow smaller than previous results \citep{Amati:2002,Amati:2003,Amati:2006}. One possible reason for the discrepancy may be that our sample is purely obtained from the Swift catalog, while previous works collected sample from other catalog (or the combination of various catalogs). For example, \citet{Amati:2006} calculated the $E_p-E_{\rm iso}$ relation of 41 GRBs from the combination of BATSE, BeppoSAX, HETE-2, Konus and Swift catalogs, and found a slope of $\sim 2$. In the $E_p-E_{\rm iso}$ plot, the Swift subsample seems to fall near the best-fitting line of the full sample. Therefore, he claimed that the Swift subsample follows the same $E_p-E_{\rm iso}$ relation of the full sample. However,  \citet{Amati:2006} has not calculated the slope parameter for the Swift subsample alone. In fact, if we only consider the 9 Swift GRBs corrected in table 1 of Amati's paper, we get a slope of $\sim 1.26$, which is much flatter than that of the full sample. For the pure Swift GRBs collected in Table 13 of \citet{Sakamoto:2011}, we get a slope of $\sim 1.47$, still much smaller than the originally proposed value $\sim 2$ \citep{Amati:2002}. This may imply that the Swift GRBs have a flatter $E_p-E_{\rm iso}$ slope than other GRBs. In the bottle two panels of Figure \ref{fig:correlations}, we plot the $1\sigma$, $2\sigma$ and $3\sigma$ contours in the $(a,b)$ plane. Red contours represent the CPL case and black contours represent the Band case. The best-fitting values are denoted by dots. It seems that both $E_p-E_{\rm iso}$ relation and $E_p-L$ relation in the Band cases are differs from that in the CPL case at about $2\sigma$ C.L. In the last column of Table \ref{tab:parameters4}, we also list the Pearson's linear correlation coefficient. We can see that the correlations in the CPL case are as tight as that in the Band case. However, both correlations are not as tight as previously claimed. The intrinsic scatters of both luminosity correlations are comparable to the results of \citet{Wang:2011}.

\begin{figure}
\centering
 \includegraphics[width=0.5\textwidth,height=12cm]{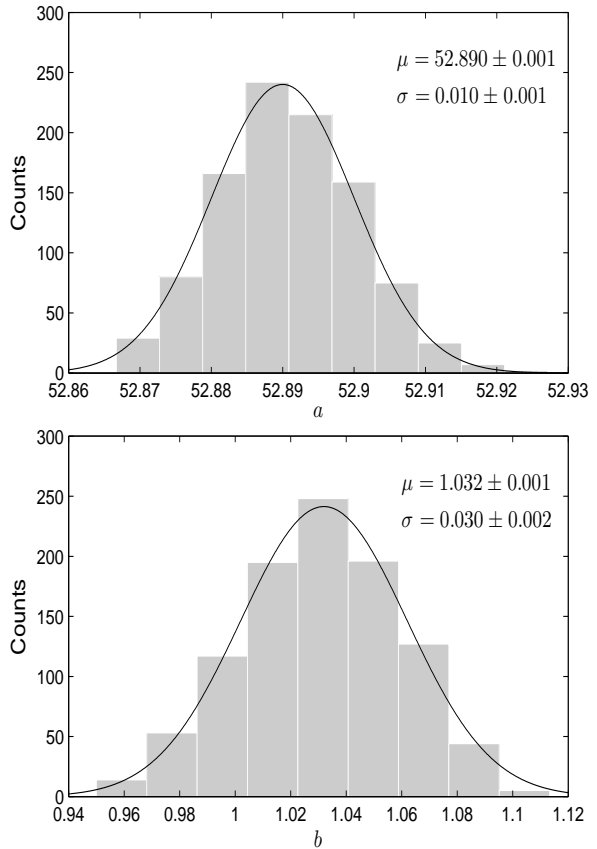}
 \caption{\small{The histograms of $a$ and $b$ in 1000 simulations. The solid curves are the best-fitting results to Gauss function.}}\label{fig:hist_ab}
\end{figure}


\begin{table*}
\centering
\caption{\small{The properties of Swift-BAT GRBs observed by Konus-Wind or Fermi-GBM. Column (1): the GRB identifier. Column (2): the redshift. Column (3): the observed photon fluence in unit of $10^{-7}$ erg cm$^{-2}$. Column (4): the low-energy photon index. Column (5): the high-energy photon index.  Column (6): the observed peak energy in unit of keV. Column (7): the energy band in which the photon fluence is calculated, in unit of keV. Column (8): the isotropic equivalent energy in unit of $10^{52}$ ergs. Column (9): the detector by which the GRB was observed, KON for Konus-Wind and GBM for Fermi-GBM. Column (10): the GCN Circular numbers. All the errors in this table are of 90\% C.L.}}\label{tab:parameters5}
\begin{tabular}{llllllllll}
\hline
(1)	&	(2)	&	(3)	&	(4)	&	(5)	&	(6)	&	(7)	&	(8)	&	(9)	&	(10)	\\
GRBs	&	$z$	&	$S$	&	$\alpha$	&	$\beta$	&	$E_p$	&	$E_{\rm min}-E_{\rm max}$	&	$E_{\rm iso}$	&	Detector	&	GCN Circ	\\
\hline
141220A	&	1.3195	&	$52.3_{-1.5}^{+1.5}$	&	$-0.80_{-0.05}^{+0.05}$	&	...	&	$180_{-9}^{+9}$	&	$8-1000$	&	$2.48_{-0.07}^{+0.07}$	&	GBM	&	17205	\\
140206A	&	2.73	&	$147_{-3}^{+3}$	&	$-0.20_{-0.10}^{+0.10}$	&	$-2.40_{-0.10}^{+0.10}$	&	$120_{-6}^{+6}$	&	$10-1000$	&	$29.14_{-0.59}^{+0.59}$	&	GBM	&	15796	\\
130925A	&	0.347	&	$2.83_{-0.06}^{+0.06}$	&	$-1.50_{-0.05}^{+0.05}$	&	...	&	$107_{-3}^{+3}$	&	$10-1000$	&	$0.01_{-0.01}^{+0.01}$	&	GBM	&	15261	\\
130701A	&	1.155	&	$58_{-2}^{+2}$	&	$-1.10_{-0.10}^{+0.10}$	&	...	&	$89_{-4}^{+4}$	&	$20-1200$	&	$2.64_{-0.09}^{+0.09}$	&	KON	&	14958	\\
120922A	&	3.1	&	$65_{-4}^{+4}$	&	$-1.60_{-0.70}^{+0.70}$	&	$-2.30_{-0.10}^{+0.10}$	&	$37.7_{-3.5}^{+3.5}$	&	$10-1000$	&	$21.59_{-1.33}^{+1.33}$	&	GBM	&	13809	\\
120326A	&	1.798	&	$35.39_{-1.74}^{+1.74}$	&	$-0.98_{-0.14}^{+0.14}$	&	$-2.53_{-0.15}^{+0.15}$	&	$46.5_{-3.7}^{+3.7}$	&	$10-1000$	&	$3.62_{-0.18}^{+0.18}$	&	GBM	&	13145	\\
110715A	&	0.82	&	$230_{-20}^{+20}$	&	$-1.23_{-0.08}^{+0.09}$	&	$-2.70_{-0.50}^{+0.20}$	&	$120_{-11}^{+12}$	&	$20-10000$	&	$4.92_{-0.43}^{+0.43}$	&	KON	&	12166	\\
110422A	&	1.77	&	$856_{-2}^{+2}$	&	$-0.65_{-0.06}^{+0.06}$	&	$-2.96_{-0.19}^{+0.14}$	&	$152_{-5}^{+5}$	&	$20-2000$	&	$74.59_{-0.17}^{+0.17}$	&	KON	&	11971	\\
100816A	&	0.8034	&	$33_{-4}^{+4}$	&	$-1.00_{-0.30}^{+0.40}$	&	...	&	$148_{-26}^{+41}$	&	$20-2000$	&	$0.65_{-0.08}^{+0.08}$	&	KON	&	11127	\\
090926B	&	1.24	&	$87_{-3}^{+3}$	&	$-0.13_{-0.06}^{+0.06}$	&	...	&	$91_{-2}^{+2}$	&	$10-1000$	&	$3.65_{-0.13}^{+0.13}$	&	GBM	&	9957	\\
090618	&	0.54	&	$2700_{-60}^{+60}$	&	$-1.26_{-0.02}^{+0.06}$	&	$-2.50_{-0.33}^{+0.15}$	&	$155.5_{-10.5}^{+11.1}$	&	$8-1000$	&	$25.87_{-0.57}^{+0.57}$	&	GBM	&	9535	\\
090424	&	0.544	&	$520_{-10}^{+10}$	&	$-0.90_{-0.02}^{+0.02}$	&	$-2.90_{-0.10}^{+0.10}$	&	$177_{-3}^{+3}$	&	$8-1000$	&	$4.55_{-0.09}^{+0.09}$	&	GBM	&	9230	\\
090423	&	8	&	$11_{-3}^{+3}$	&	$-0.77_{-0.35}^{+0.35}$	&	...	&	$82_{-15}^{+15}$	&	$8-1000$	&	$10.61_{-2.89}^{+2.89}$	&	GBM	&	9229	\\
081222	&	2.77	&	$132_{-40}^{+48}$	&	$-0.67_{-0.33}^{+0.39}$	&	$-2.35_{-1.25}^{+0.30}$	&	$165_{-29}^{+47}$	&	$20-1000$	&	$28.67_{-8.69}^{+10.43}$	&	KON	&	8721	\\
081221	&	2.26	&	$228_{-7}^{+7}$	&	$-0.91_{-0.14}^{+0.15}$	&	...	&	$83_{-4}^{+4}$	&	$20-1000$	&	$35.11_{-1.08}^{+1.08}$	&	KON	&	8694	\\
081121	&	2.512	&	$179_{-31}^{+37}$	&	$-0.77_{-0.14}^{+0.15}$	&	$-2.51_{-0.66}^{+0.31}$	&	$248_{-32}^{+38}$	&	$20-7000$	&	$25.78_{-4.46}^{+5.33}$	&	KON	&	8548	\\
080916A	&	0.689	&	$55.4_{-6.6}^{+7.9}$	&	$-1.00_{-0.32}^{+0.38}$	&	...	&	$129_{-21}^{+33}$	&	$20-1000$	&	$0.81_{-0.10}^{+0.12}$	&	KON	&	8259	\\
080913	&	6.44	&	$8.5_{-2.2}^{+6.0}$	&	$-0.89_{-0.46}^{+0.65}$	&	...	&	$131_{-48}^{+225}$	&	$15-1000$	&	$6.20_{-1.60}^{+4.37}$	&	KON	&	8256	\\
080605	&	1.6398	&	$302_{-12}^{+13}$	&	$-1.03_{-0.07}^{+0.07}$	&	...	&	$252_{-17}^{+20}$	&	$20-2000$	&	$22.82_{-0.91}^{+0.98}$	&	KON	&	7854	\\
080603B	&	2.69	&	$45.0_{-9.0}^{+15.3}$	&	$-1.23_{-0.54}^{+0.75}$	&	...	&	$102_{-28}^{+119}$	&	$20-1000$	&	$9.81_{-1.96}^{+3.33}$	&	KON	&	7812	\\
071010B	&	0.947	&	$47.8_{-31.2}^{+9.5}$	&	$-1.25_{-0.49}^{+0.74}$	&	$-2.65_{-0.49}^{+0.29}$	&	$52_{-14}^{+10}$	&	$20-1000$	&	$1.76_{-1.15}^{+0.35}$	&	KON	&	6879	\\
070521	&	0.553	&	$181_{-31}^{+6}$	&	$-0.93_{-0.12}^{+0.12}$	&	...	&	$222_{-21}^{+27}$	&	$20-1000$	&	$1.58_{-0.27}^{+0.05}$	&	KON	&	6459	\\
070508	&	0.82	&	$397_{-23}^{+7}$	&	$-0.81_{-0.07}^{+0.07}$	&	...	&	$188_{-8}^{+8}$	&	$20-1000$	&	$7.72_{-0.45}^{+0.14}$	&	KON	&	6403	\\
050525A	&	0.606	&	$784_{-6}^{+6}$	&	$-1.10_{-0.05}^{+0.05}$	&	...	&	$84.1_{-1.7}^{+1.7}$	&	$20-1000$	&	$9.81_{-0.08}^{+0.08}$	&	KON	&	3474	\\
\hline
\end{tabular}
\end{table*}

Among the 44 Swift-BAT GRBs considered here, at least 24 of them are also observed by Konus-Wind \citep{Aptekar:1995} or Fermi-GBM \citep{Meegan:2009}. These two detectors can observe GRBs in a much wider energy band (10 keV $-$ 10 MeV for Konus-Wind and 8 keV $-$ 40 MeV for Fermi-GBM) than Swift-BAT, making the detailed analysis of spectra possible. We list the spectral properties of these 24 GRBs in Table \ref{tab:parameters5}. The spectra of 10 GRBs can be fitted to the Band function, so that the high-energy index $\beta$ can be well constrained. For the remain 14 GRBs, the spectra can only be fitted to CPL, and $\beta$ still cann't be constrained. In column (8) of Table \ref{tab:parameters5}, we also list the isotropic-equivalent energy $E_{\rm iso}$. For GRBs which have well measured $\beta$, $E_{\rm iso}$ is calculated from the Band model. Otherwise, $E_{\rm iso}$ is calculated from the CPL model. With these 24 GRBs, we calculate the Amati relation and obtain parameters $a=52.749\pm 0.143$, $b=1.285\pm 0.538$, $\sigma_{\rm int}=0.701\pm 0.101$. Due to the narrow coverage of $E_p$ and the small sample, the intrinsic scatter is large and the slope parameter couldn't be well constrained. Nevertheless, the slope still seems to be flatter than previous findings, but consistent with our results. It is not clearly understood why Swift GRBs have flatter slope. One reason may be that the peak energies  measured by Swift/BAT have a relatively narrow distribution (e.g., most of $E_{p,i}$'s are between 50 and 500 keV), while the samples investigated previously cover at least 3 orders of magnitude in $E_{p,i}$ (from a few keV to a few thousands keV).

To test the sensitivity of luminosity correlations on the high-energy power-law index $\beta$, we implement the Monte Carlo simulations. We recalculate the Amati relation in the case that GRB spectra are modeled by the Band function. In this time, $\beta$ is no longer fixed at $-2.2$, but it is a random number generated from the Gauss distribution with average value $-2.2$ and standard deviation $0.4$\footnote{We constrain $\beta$ to be smaller than $-2$, such that $E_p$ is well defined. This can be done by setting $\beta$ to its symmetrical value for any $\beta\geq -2$, i.e., using $\beta'=-4.4-\beta$ instead of $\beta$. The price is that the distribution of $\beta$ is not strictly Gaussian.}. We simulate 1000 times and derive the intercept and slope parameters $(a,b)$ of Amati relation in each simulation. The histograms of $a$ and $b$ in 1000 simulations are plotted in Figure \ref{fig:hist_ab}.
The distribution of both $a$ and $b$ can be well fitted by Gauss function, with average values and standard deviations $\mu_a=52.890\pm 0.001$, $\sigma_a=0.010\pm 0.001$ ($95\%$ C.L.), and $\mu_b=1.032\pm 0.010$, $\sigma_b=0.030\pm 0.002$ ($95\%$ C.L.), respectively. This is consistent with the Amati relation in both the CPL and Band cases (see Table \ref{tab:parameters4}). The small standard deviations of $a$ and $b$ imply that the Amati relation is insensitive to $\beta$. Since the Yonetoku relation in the CPL cases does not much differ from that in the Band case, we can expect that it is also insensitive to $\beta$.

\section{Distance calibration and GRB Hubble diagram}\label{sec:hubble}

\begin{figure*}
  \centering
 \includegraphics[width=0.8\textwidth]{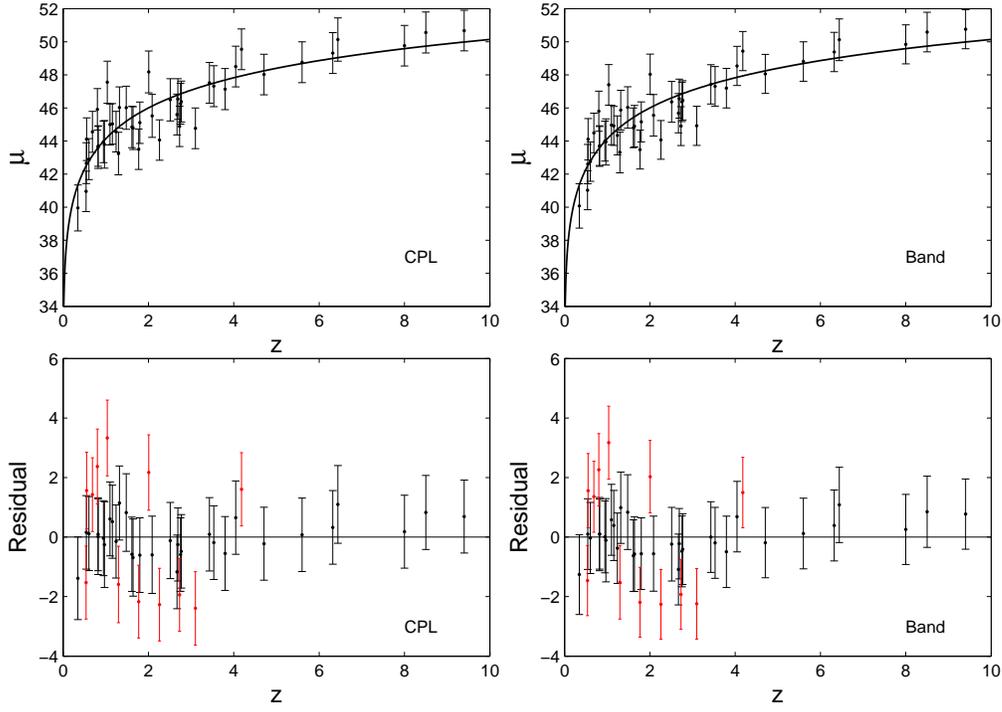}
 \caption{\small{The GRB Hubble diagram and residuals of distance moduli, calibrated using the Amati relation. Left two panels: in the case of CPL spectrum. Right two panels: in the case of Band spectrum. In the top two panels, the black curve is the theoretical result of concordance $\Lambda$CDM model ($\Omega_M=0.28$, $\Omega_{\Lambda}=0.72$, $H_0=70~{\rm km}~{\rm s}^{-1}~{\rm Mpc}^{-1}$). In the bottle two panels, the black (red) error bars represent that the calibrated distance moduli are consistent (inconsistent) with the $\Lambda$CDM model within $1\sigma$ uncertainty.}}\label{fig:hubble1}
\end{figure*}

\begin{figure*}
  \centering
 \includegraphics[width=0.8\textwidth]{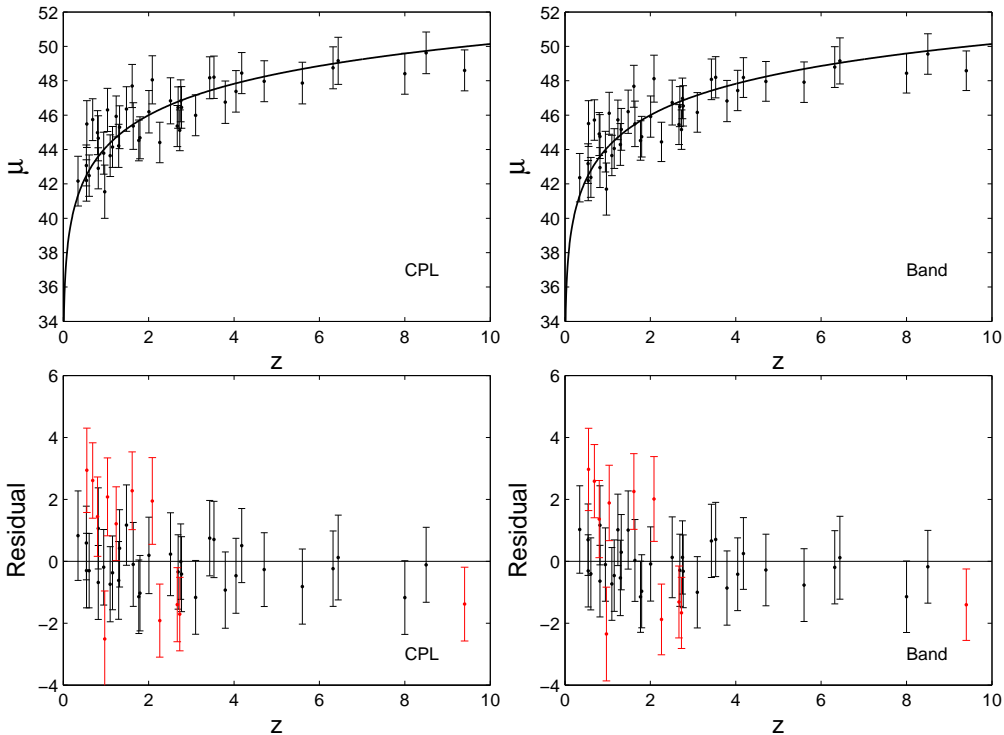}
 \caption{\small{The same to Figure \ref{fig:hubble1}, except that the distance moduli are calibrated using the Yonetoku relation.}}\label{fig:hubble2}
\end{figure*}

Both the Amati relation and Yonetoku relation have been widely used to calibrate the distance of GRBs. Here we take the Amati relation as an example. The calibrating procedures are as follows: first, derive the best-fitting parameters ($a,b$) of Amati relation as discussed in the last section. Second, calculate $E_{\rm iso}$ of each GRB through Eq.(\ref{eq:amati}), by fixing ($a,b$) at the best-fitting values. Then the luminosity distance of each GRB can be obtained through Eq.(\ref{eq:iso_energy}). Finally, convert the luminosity distance to distance modulus, i.e.,
\begin{equation}\label{eq:dis_modulus}
\mu(z)=5\log\frac{d_L(z)}{\rm{Mpc}}+25.
\end{equation}
The uncertainty of distance modulus is calculated using the formulae of error propagation \citep{Schaefer:2007},
\begin{equation}
  \sigma_{\mu}^2=\left(\frac{5}{2\ln 10}\right)^2\left[(\ln 10)^2\sigma_{\log E_{\rm iso}}^2 + \frac{\sigma_{S_{\rm bolo}}^2}{S_{\rm bolo}^2}\right],
\end{equation}
where
\begin{equation}
  \sigma_{\log E_{\rm iso}}^2=\sigma_a^2 + \left(\sigma_b\log\frac{E_{p,i}}{300~{\rm keV}}\right)^2 + \left(\frac{b}{\ln 10}\frac{\sigma_{E_{p,i}}}{E_{p,i}}\right)^2 + \sigma_{\rm int}^2.
\end{equation}
We can also calibrate GRBs using the Yonetoku relation in a very similar way.

We must point out that the calibrating method mentioned above has the circularity problem and is model dependent, because we have used the concordance cosmological model with fiducial parameters in the calibration. Our paper is not aimed at solving this problem, but we are rather interested in studying the effect of GRB spectra on the distance calibration. Some model-independent methods free of circularity problem have been extensively discussed \citep{Liang:2008a,Liang:2008b,Wei:2009,Wei:2010,Liu:2014}.

The distance moduli calibrated using the Amati relation and Yonetoku relation are listed in columns $(6)-(7)$ in Table \ref{tab:parameters2} and Table \ref{tab:parameters3}, respectively. We can see that the distance moduli only slightly depend on the choice of GRB spectra. The difference between distance moduli in the two spectra cases (Band and CPL) is much smaller than the uncertainties of distance moduli, i,e, $|\mu^{\rm Band}-\mu^{\rm CPL}|\ll \sigma_{\mu}$. In Figure \ref{fig:hubble1}, we plot the GRB Hubble diagram calibrated using Amati relation. The top-left and top-right panels are the resulting Hubble diagram in the CPL and Band cases, respectively. The solid curves are the theoretical results of concordance $\Lambda$CDM model with fiducial parameters $H_0=70~{\rm km}~{\rm s}^{-1}~{\rm Mpc}^{-1}$, $\Omega_M=0.28$ and $\Omega_{\Lambda}=0.72$. In the bottle two panels of Figure \ref{fig:hubble1}, we plot the residual of distance moduli with respect to the theoretical curve. The error bar represents $1\sigma$ uncertainty. GRBs consistent (inconsistent) with theoretical curve within $1\sigma$ uncertainty are denoted by black (red). Due to the large intrinsic scatter of Amati relation, 27\% (12 out of 44) GRBs do not fall onto the theoretical curve within $1\sigma$ uncertainty. These GRBs all have low redshift ($z\lesssim 4$). GRBs inconsistent with theoretical curve in the CPL case are still inconsistent in the Band case. The uncertainty of distance modulus mainly ($\gtrsim 90\%$) propagates from the intrinsic scatter. A much tighter correlation with smaller intrinsic scatter is necessary to reduce $\sigma_{\mu}$.

We also reconstruct the GRB Hubble diagram using the Yonetoku relation and plot it in Figure \ref{fig:hubble2}. Compared to the Hubble diagram reconstructed from Amati relation, very similar conclusion can be draw. Therefore, we conclude that the GRB Hubble diagram almost does not depend on the choice of GRB spectra.

\section{Summary}\label{sec:conclusions}

In this paper, we have investigated the effect of GRB spectra on the luminosity correlations and GRB Hubble diagram. Most previous works calculated the isotropic equivalent energy and peak luminosity by assuming that GRB spectra are modeled by the Band function. Due to the narrow energy coverage of detectors, the high-energy power-law index of Band function usually can't be constrained. Using 44 Swift GRBs with well determined peak energy, we calculated two luminosity correlations (Amati relation and Yonetoku relation) in the cases that GRB spectra are modeled by the Band function and CPL, respectively. It is found that both Amati relation and Yonetoku relation only moderately depend on the choice of GRB spectra. Monte Carlo simulations show that the Amati relation is insensitive to the high-energy power-law index of Band function. We further calibrated the distance of GRBs using these two luminosity correlations and reconstructed the GRB Hubble diagram. We found that different GRB spectra only cause the difference of distance modulus at the order of 0.1 mag, while the uncertainty of distance modulus calibrated using luminosity correlations are usually larger than 1.0 mag. Therefore, we may conclude that the resulting GRB Hubble diagram is insensitive to the choice of GRB spectra.

\section*{Acknowledgements}
We are grateful to J. Li, H. Ma and L. Tang for useful discussions. X. Li has been supported by the National Natural Science Fund of China (NSFC) (Grant No. 11305181 and 11547305) and the Open Project Program of State Key Laboratory of Theoretical Physics, Institute of Theoretical Physics, Chinese Academy of Sciences, China (No. Y5KF181CJ1). Z. Chang has been funded by the NSFC under Grant No. 11375203.

\label{lastpage}


\begin{thebibliography}{}

\bibitem[\protect\citeauthoryear{Amati et al.}{2002}]{Amati:2002}Amati L., et al., 2002, A\&A, 390, 81

\bibitem[\protect\citeauthoryear{Amati}{2008}]{Amati:2008}Amati L., et al., 2008, MNRAS, 391, 577

\bibitem[\protect\citeauthoryear{Amati}{2003}]{Amati:2003}Amati L., 2003, Chin. J. Astron. Astrophys. Supp., 3, 455

\bibitem[\protect\citeauthoryear{Amati}{2006}]{Amati:2006}Amati L., 2006, MNRAS, 372, 233

\bibitem[\protect\citeauthoryear{Amati}{2010}]{Amati:2010}Amati L., 2010, arXiv:1002.2232

\bibitem[\protect\citeauthoryear{Amati, Frontera \& Guidorzi}{2009}]{Amati:2009}Amati L., Frontera F., Guidorzi C., 2009, Astron. Asrophys., 508, 173

\bibitem[\protect\citeauthoryear{Aptekar et al.}{1995}]{Aptekar:1995}Aptekar R. L., et al., 1995, Space Sci. Rev., 71, 265

\bibitem[\protect\citeauthoryear{Band et al.}{1993}]{Band:1993}Band D., et al., 1993, ApJ, 413, 281

\bibitem[\protect\citeauthoryear{Basilakos \& Perivolaropoulos}{2008}]{Basilakos:2008}Basilakos S., Perivolaropoulos L., 2008, MNRAS, 391, 411

\bibitem[\protect\citeauthoryear{Bernardini et al.}{2012}]{Bernardini:2012}Bernardini M. G., Margutti R., Zaninoni E., Chincarini G., 2012, MNRAS, 425, 1199

\bibitem[\protect\citeauthoryear{Cabrera et al.}{2007}]{Cabrera:2007}Cabrera J. I., et al., 2007, MNRAS, 382, 342

\bibitem[\protect\citeauthoryear{Cucchiara et al.}{2011}]{Cucchiara:2011}Cucchiara A., et al., 2011, ApJ, 736, 7

\bibitem[\protect\citeauthoryear{Dai, Liang \& Xu}{2004}]{Dai:2004tq}Dai Z. G., Liang E. W., Xu D., 2004, Astrophys. J., 612, L101

\bibitem[\protect\citeauthoryear{Dainotti et al.}{2013}]{Dainotti:2013}Dainotti M. G., Cardone V. F., Piedipalumbo E., Capozziello S., 2013, MNRAS, 436, 82

\bibitem[\protect\citeauthoryear{D'Agostini}{2005}]{DAgostini:2005}D'Agostini G., 2005, arXiv:physics/0511182 [physics.data-an]

\bibitem[\protect\citeauthoryear{Fenimore \& Ramirez-Ruiz}{2000}]{Fenimore:2000}Fenimore E. E., Ramirez-Ruiz E., 2000, arXiv: astro-ph/0004176

\bibitem[\protect\citeauthoryear{Firmani et al.}{2005}]{Firmani:2005}Firmani C., Ghisellini G., Ghirlanda G., Avila-Reese V., 2005, MNRAS, 360, L1

\bibitem[\protect\citeauthoryear{Firmani et al.}{2006}]{Firmani:2006b}Firmani C., Ghisellini G., Avila-Reese V., Ghirlanda G., 2006, MNRAS, 370, 185

\bibitem[\protect\citeauthoryear{Fishman \& Meegan}{1995}]{Fishman:1995}Fishman G., Meegan C., 1995, Annu. Rev. Astron. Astrophys., 33, 415

\bibitem[\protect\citeauthoryear{Ghirlanda, Ghisellini \& Lazzati}{2004}]{Ghirlanda:2004b}Ghirlanda G., Ghisellini G., Lazzati D., 2004, ApJ, 616, 331

\bibitem[\protect\citeauthoryear{Ghirlanda et al.}{2004}]{Ghirlanda:2004a}Ghirlanda G., Ghisellini G., Lazzati D., Firmani C., 2004, ApJ, 613, L13

\bibitem[\protect\citeauthoryear{Kouveliotou et al.}{1993}]{Kouveliotou:1993}Kouveliotou C., et al., 1993, ApJ, 413, L101

\bibitem[\protect\citeauthoryear{Kumar \& Zhang}{2015}]{Kumar:2015}Kumar P., Zhang B., 2015, Phys. Rep., 561, 1

\bibitem[\protect\citeauthoryear{Li}{2007}]{Li:2007}Li L. X., 2007, MNRAS, 379, L55

\bibitem[\protect\citeauthoryear{Liang et al.}{2015}]{Liang:2015dua}Liang E.~W., Lin T.~T., LV J., Lu R., Zhang J., Zhang B., 2015, arXiv:1505.03660

\bibitem[\protect\citeauthoryear{Liang et al.}{2008}]{Liang:2008a}Liang N., Xiao W. K., Liu Y., Zhang S. N., 2008, ApJ, 685, 354

\bibitem[\protect\citeauthoryear{Liang \& Zhang}{2005}]{Liang:2005}Liang E. W., Zhang B., 2005, ApJ, 633, 611

\bibitem[\protect\citeauthoryear{Liang \& Zhang}{2008}]{Liang:2008b}Liang N., Zhang S. N., 2008, AIP Conf. Proc., 1065, 367

\bibitem[\protect\citeauthoryear{Lin et al.}{2015}]{Lin:2015a}Lin H.-N., Li X., Wang S., Chang Z., 2015, MNRAS, 453, 128

\bibitem[\protect\citeauthoryear{Lin, Li \& Chang}{2015}]{Lin:2015b}Lin H.-N., Li X., Chang Z., 2015, arXiv:1507.06662, accepted by MNRAS

\bibitem[\protect\citeauthoryear{Liu \& Wei}{2014}]{Liu:2014}Liu J., Wei H., 2014, arXiv:1410.3960

\bibitem[\protect\citeauthoryear{Margutti et al.}{2013}]{Margutti:2013}Margutti R., et al., 2013, MNRAS, 428, 729

\bibitem[\protect\citeauthoryear{Meegan et al.}{2009}]{Meegan:2009}Meegan C. A., Lichti G., Bhat P. N., et al., 2009, ApJ, 702, 791

\bibitem[\protect\citeauthoryear{M\'{e}sz\'{a}ros}{2006}]{Meszaros:2006}M\'{e}sz\'{a}ros P., 2006, Rep. Prog. Phys., 69, 2259

\bibitem[\protect\citeauthoryear{Norris, Marani \& Bonnell}{2000}]{Norris:2000}Norris J. P., Marani G. F., Bonnell J. T., 2000, ApJ, 534, 248

\bibitem[\protect\citeauthoryear{Paciesas et al.}{1999}]{Paciesas:1999}Paciesas W. S., et al., 1999, ApJ, 122, S465

\bibitem[\protect\citeauthoryear{Piran}{1999}]{Piran:1999}Piran T., 1999, Phys. Rep., 314, 575

\bibitem[\protect\citeauthoryear{Preece et al.}{2000}]{Preece:2000}Preece R. D., et al., 2000, ApJ, 126, S19

\bibitem[\protect\citeauthoryear{Sakamoto et al.}{2011}]{Sakamoto:2011}Sakamoto T., et al., 2011, ApJ, 195, S2

\bibitem[\protect\citeauthoryear{Schaefer}{2003}]{Schaefer:2003}Schaefer B. E., 2003, ApJ, 583, L67

\bibitem[\protect\citeauthoryear{Schaefer}{2007}]{Schaefer:2007}Schaefer B. E., 2007, ApJ, 660, 16

\bibitem[\protect\citeauthoryear{Wang, Qi \& Dai}{2011}]{Wang:2011}Wang F. Y., Qi S., Dai Z. G., 2011, MNRAS, 415, 3423

\bibitem[\protect\citeauthoryear{Wang et al.}{2015}]{Wang:2015cya}Wang J.~S., Wang F.~Y., Cheng K.~S., Dai Z.~G., 2015, arXiv:1509.08558

\bibitem[\protect\citeauthoryear{Wei \& Zhang}{2009}]{Wei:2009}Wei H., Zhang S. N., 2009, Eur. Phys. J. C, 63, 139

\bibitem[\protect\citeauthoryear{Wei}{2010}]{Wei:2010}Wei H., 2010, J. Cosmol. Astropart. Phys., 08, 020

\bibitem[\protect\citeauthoryear{Xiao \& Schaefer}{2009}]{Xiao:2009}Xiao L. M., B. E. Schaefer, 2009, ApJ, 707, 387

\bibitem[\protect\citeauthoryear{Yonetoku et al.}{2004}]{Yonetoku:2004}Yonetoku D., Murakami T., Nakamura T., Yamazaki R., Inoue A. K., Ioka K., 2004, ApJ, 609, 935

%
%
%
%
%
%
%
%
%
%
%
%
%
%
%
%
%
%
%
%
%
%
%
%
%
%
%
%
%
%
%
\end{thebibliography}
\end{document}